%% file: main.tex
\documentclass[%reprint,
superscriptaddress,
 amsmath,amssymb,
 aps,
prl,
onecolumn,
]{revtex4-2}

\newcommand{\ca}{$^{40}$Ca$^+$}
\newcommand{\be}{$^{9}$Be$^+$}
\usepackage{braket}
\usepackage{graphicx}
\usepackage[colorlinks=true, allcolors=blue]{hyperref}
\usepackage{gensymb}
\usepackage{derivative}
\usepackage{upgreek}

\usepackage{float}  
\usepackage{xcolor}
\usepackage{footmisc}
\usepackage{siunitx}

\usepackage{calrsfs}
\usepackage{mathrsfs}

\usepackage{braket}
\usepackage{graphicx}
\usepackage{booktabs} 
\usepackage[colorlinks=true, allcolors=blue]{hyperref}
\usepackage{upgreek}
\newsavebox\mcFcontent
\savebox\mcFcontent{$\mathcal{F}$}

\begin{document}

\title{Quantum Vector Signal Analyzer: Wideband Electric Field Sensing via Motional Raman Transitions}
\author{Hao Wu}
\thanks{hao.wu@physics.ucla.edu}
\author{Grant D. Mitts}
\author{Clayton Z.C. Ho}
\author{Joshua A. Rabinowitz}
\author{Eric R. Hudson}
\affiliation{Department of Physics and Astronomy, University of California Los Angeles, Los Angeles, CA, USA}
\affiliation{Challenge Institute for Quantum Computation, University of California Los Angeles, Los Angeles, CA, USA}
\affiliation{Center for Quantum Science and Engineering, University of California Los Angeles, Los Angeles, CA, USA}

\date{\today} 
\maketitle

\textbf{Ultrasensitive detection of the frequency, phase, and amplitude of radio frequency (RF) electric fields is central to a variety of important applications, including radio communication, cosmology~\cite{Spitler2016},  dark matter searches~\cite{dark2003}, and high-fidelity qubit control~\cite{Jerger2019}. 
Quantum harmonic oscillator systems, especially trapped ions, have been used with several quantum sensing techniques~\cite{Wolf2019,McCormick2019,Burd2019,Keller2021} to achieve electric field sensing with state-of-the-art sensitivity~\cite{Gilmore2021} and nanometer spatial resolution~\cite{Valdis2018}.
However, these systems are limited to a narrow frequency range centered around either the motional frequency of the trapped ion oscillator or the frequency of an optical transition in the ion; often these techniques are not sensitive to the RF phase.
Here, we propose and demonstrate a procedure that unlocks the extreme sensitivity of a quantum harmonic oscillator to allow high precision \emph{wideband} detection of the frequency, phase, and amplitude of an unknown electric field.
Specifically, we use motional Raman transitions in a single trapped ion, cooled near its motional ground state to realize state of the art sensitivities to frequency, phase, and amplitude, and show the technique works over a frequency range that is $>800\times$ larger than previous techniques. 
Further, this technique is shown to be compatible with both quantum amplification via squeezing and measurement in the Fock basis, allowing a demonstration of performance 3.4(20) dB below the standard quantum limit and the potential for several orders of magnitude improvement in sensitivity with moderate upgrades. 
In addition to providing an attractive platform for quantum sensing of small fields, this technique also allows \emph{in situ} calibration of qubit control lines in systems using quantum harmonic oscillators, as well as transduction of external, non-resonant drives into oscillator excitation.
Additionally, this approach can be extended to other quantum harmonic oscillator systems, such as a superconducting qubit-resonator system.
} \newline

Well-controlled quantum systems offer quantum-limited metrology capabilities~\cite{Pezze2018,Degen2017}. 
Quantum harmonic oscillators (QHOs) controlled at the single quanta level serve as an excellent platform for high-sensitivity detection of radio frequency (RF) electric fields.
For a single QHO mode, it has been shown that non-classical Fock states provide optimal precision and sensitivity for measurements of displacements with unknown phase~\cite{Wolf2019}. 
Such states have been prepared in trapped ions to measure the amplitude~\cite{Wolf2019} and frequency~\cite{Wolf2019,McCormick2019} of an oscillator below the standard quantum limit (SQL).
Additionally, a Fock state with up to $N = 100$ quanta in a superconducting microwave cavity was used to demonstrate displacement and oscillator frequency sensitivity approaching the Heisenberg limit~\cite{deng2024}. 

By employing squeezing to achieve ‘quantum amplification’, the sensitivity of a QHO to resonant forces has been enhanced by approximately an order of magnitude~\cite{Burd2019}.
Additionally, measurements utilizing the entanglement of collective spin states of trapped ions with a shared motional mode have shown sensitivity to displacements approximately 9 dB below the SQL~\cite{Gilmore2021}. 
Similarly, NOON states have been used to realize frequency sensitivities near the Heisenberg limit~\cite{Zhang2018}, while important work with motional cat states~\cite{Milne2019} and amplitude-modulated displacements of Fock states~\cite{Keller2021} have demonstrated the ability to measure the noise spectrum of an oscillator from near DC to the natural frequency of the oscillator.
Meanwhile, to achieve highly sensitive off-resonance measurements, spin-dependent optical dipole forces (ODF) have been used to map the amplitude of the motion oscillating at frequencies offset by $\sim$MHz from the oscillator's natural frequency onto the internal spin, allowing the detection of displacements 40 times smaller than the zero-point fluctuations of the oscillator~\cite{Gilmore2017}. Achieving a more stable relative phase between the displacement and ODF  could further enhance the sensitivity by an additional order of magnitude\cite{affolter2020}.

Despite these demonstrations of state-of-the-art sensitivities~\cite{Gilmore2021}, comparable to those achieved on other platforms such as NV centers~\cite{NV_review,Qiu2022} and Rydberg atoms~\cite{Liu2022,Holloway2022}, challenges remain.
First, these techniques are often only sensitive to the amplitude of the oscillator's displacement, and are therefore unable to sense the phase of the electric field. 
Techniques that do have phase sensitivity, such as the phase-sensitive red sideband (PSRSB)~\cite{Burd2019,Hempel2013}, phase-dependent energy shift induced by spin-dependent ODF~\cite{affolter2020}, and quantum amplification methods~\cite{Burd2019}, require an RF electric field frequency that is near either frequency of an optical transition inside the ion or the ion motional frequency, respectively.
Further, implementations of either the PSRSB method or phase-dependent energy shift with optical fields is compromised by the difficult-to-control optical phase noise~\cite{Day2022}, though this is overcome on dedicated platforms able to generate microwave-field gradients~\cite{burd2023experimental}, while the increased decoherence associated with quantum amplification significantly limits interrogation times and therefore worsens the achievable sensitivities~\cite{Burd2019}.
Second, even for measurements of only the amplitude and frequency of the electric field, the accessible frequency range is limited to near the natural frequency of the oscillator, with achieved bandwidths of only several MHz.

Fundamentally, these restrictions, which have significantly limited the application of QHOs for electric field sensing, arise due to the narrowband response of QHOs.
Here, we show that a phase-sensitive, heterodyne detection scheme can be constructed from the interference of multiple motional Raman transitions that provides the ability to ``continuously" sense the amplitude, frequency, and phase of an unknown field across a wide band with high precision. 
In what follows, we describe the interaction responsible for and the operation of a wideband quantum vector signal analyzer (QVSA) based on a single trapped ion.
We benchmark the method by measuring the known transfer function of a commercial filter and demonstrate \emph{in situ} direct calibration of a qubit control line. 
Further, we characterize the sensitivity and precision of the QVSA at 82~MHz for use in future molecular ion qubit gates~\cite{EGGs} and measure an amplitude sensitivity (without the use of a pre-amplifier) of \SI{51(10)}{\micro V/\sqrt{\textup{Hz}}}, corresponding to an electric field sensitivity of 66(13)~mV$\textrm{m}^{-1}$/$\sqrt{\textrm{Hz}}$ with a minimum detectable voltage of~\SI{3.8(8)}{\micro V} (4.9(1.0)~mV/m); a frequency sensitivity of 39(9)~Hz/$\sqrt{\textup{Hz}}$ with minimum resolution 3.8(1.6)~Hz; and a phase sensitivity of 117(26)~mrad/$\sqrt{\textup{Hz}}$ with minimum phase resolution~$9.1(2.0)$~mrad. 
We also show that the technique is compatible with quantum amplification~\cite{Burd2019,Burd2021,burd2023experimental,Ge2019} and use a squeezing operation to realize a 9.0(4)~dB gain in voltage sensitivity and demonstrate an amplitude sensitivity 3.4(2.0)~dB below the SQL.
Finally, we demonstrate the technique across the low-frequency RF to the UHF microwave frequency bands (100 kHz to 1 GHz), a range primarily limited by our signal synthesis capabilities, compare the measured electric field sensitivity to other platforms, and discuss several straightforward modifications for improving the technique.
This advancement paves the way for utilizing QHOs for wideband electric field sensing.

To demonstrate this technique, we use a QHO realized via a radial motional mode at $\omega \approx 2\pi\cdot1$~MHz of a single \ca\ ion held in a cryogenically cooled (4.5~K) linear Paul trap with an ion-electrode distance $r_o$ = \SI{550}{\micro m}.
The QHO is initialized and read out via coupling through a Jaynes-Cummings Hamiltonian to an optical qubit defined on the $\{\ket{^2D_{\frac{5}{2}},m_J=-\frac{5}{2}},\ket{^2S_{\frac{1}{2}},m_J=-\frac{1}{2}}\}$ electronic states of \ca.
Doppler cooling of the \ca\ ion followed by resolved sideband cooling on the narrow optical qubit transition \cite{Roos1999,Leibfried2003} prepares the motional ground state with a measured mean phonon number of $\langle n \rangle = 0.01(1)$~(see Appendix E of the SI).

\begin{figure}
    \centering
    \includegraphics[width = 1.0\textwidth]{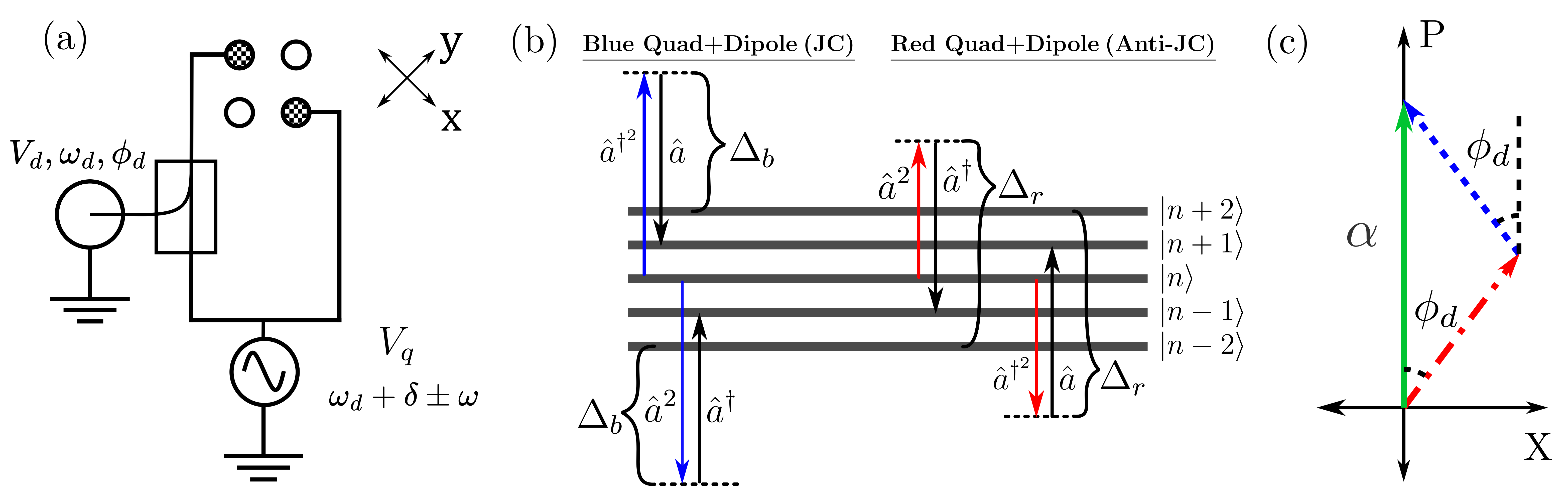}
     \caption{Implementation of the quantum vector signal analyzer (QVSA). 
    (a) Schematic of the QVSA drive coupling to the quadrupole ion trap. 
    Two quadrupole tones ($\omega_d+\delta \pm \omega$) are coupled onto the electrodes to characterize an unknown dipole tone ($\omega_d$).
    The dipole tone is easily supplied using either a BNC connector or a microwave horn.
    (b) Level diagrams for motional Raman transitions.
    The blue (red) arrow indicates the blue (red) quadrupole field, while the black arrow represents the dipole field. The combination of the blue quadrupole~($\omega_d+\delta+\omega$) and dipole fields results in displacement following the Jaynes–Cummings (JC) model, whereas the combination of the red quadrupole~($\omega_d+\delta-\omega$) and dipole fields results in displacement according to the anti-Jaynes–Cummings (Anti-JC) model.
    $\Delta_b$=$\omega_d$-$\omega$;
    $\Delta_r$=$\omega_d$+$\omega$;
    (c) Phase space diagram showing that the phase sensitivity of the QVSA displacement $\alpha$ arises from the interference between the displacements generated by the multiple Raman transitions; each transition is excited by one quadrupole and one dipole tone.
    We assume the initial phase of the quadrupole tones to be zero.}\label{fig:qvsa}
    
\end{figure}

To understand the operation of the QVSA, suppose an unknown oscillatory signal with voltage $V_d$ at frequency $\omega_d$ and phase $\phi_d$ is applied to the trap in a dipole configuration (the dipole tone) while two voltages at $V_q$ with frequencies at $\omega_d + \delta \pm \omega$ are applied to the trap in a quadrupole configuration (the quadrupole tones), where $\omega$ is the secular frequency of a trapped ion motional mode and $\delta$ is the frequency detuning as shown in Fig.~\ref{fig:qvsa}(a). 
We define the initial phase of the quadrupole tones as zero (see Appendix B of the SI).
The motion of the ion is described by the Hamiltonian:
\begin{align}\label{Hamil}
    \frac{\hat{H}}{\hbar} =  \omega \hat{a}^{\dagger} \hat{a}
    &+ \frac{\Omega_q}{2} \left( \hat{a} + \hat{a}^{\dagger} \right)^2
    \cos{ \left( \left(\omega_d + \delta+\omega\right) t \right)}  \nonumber \\
    &+\frac{\Omega_q}{2} \left( \hat{a} + \hat{a}^{\dagger} \right)^2
    \cos{ \left( \left(\omega_d + \delta-\omega\right) t \right)} \nonumber\\
    & + \Omega_d \left( \hat{a}+\hat{a}^\dagger \right) \cos{ \left( \omega_d t+\phi_d \right) }.
\end{align}
where $\Omega_q = eV_q x_o^2/(\hbar r_o^2)$, $\Omega_d = \kappa eV_d x_o/(\hbar  r_o)$, $r_o$ is the ion-electrode distance, 
$x_o = \sqrt{\hbar/(2m\omega)}$ is the zero-point wavefunction of the QHO, and $\kappa$ is a geometrical factor that depends on the motional mode. 

The resulting evolution can be understood by noticing that the dipole tone off-resonantly drives $\Delta n = \pm1$ transitions, while the quadrupole tones off-resonantly drive $\Delta n = \pm2$ transitions (see Fig.~\ref{fig:qvsa}(b)). We further assume $\omega_d \gg \omega \gg \delta$
\footnote{This assumption is made solely to simplify the analytical formula presented here.
Without this simplification, additional off-resonance displacement terms need to be included. 
However, if a sin$^2$ pulse shaping technique \cite{Zarantonello2019} is employed, the analytical formula Eq.~\eqref{eq2} remains valid without this assumption.}
, and thus for $\delta < \Omega_{q}\Omega_{d}/\left( \omega_{d} - 2 \omega \right)$ the dipole tone interacts with either quadrupole tone to drive a three-phonon Raman transition, resulting in an overall $\Delta n = \pm1$.
Thus, it is not surprising that to second order in the Magnus expansion the time evolution in the interaction picture with respect to the harmonic oscillator is approximately given by a displacement operator with displacement (see Appendix A of the SI):
\begin{align}
    \alpha = \imath  \Omega_q\Omega_d \frac{\omega}{\omega_d^2-\omega^2}\frac{\sin{(\delta t -\phi_d) + \sin{(\phi_d})}}{\delta}.
    \label{eq2}
\end{align}
As a result, the magnitude of the displacement $|\alpha|$  under this interaction depends on the amplitude $V_d$, frequency $\omega_d$, QVSA interaction time $t$, and phase $\phi_d$ of the unknown dipole tone. 

While dependence on $V_d$ and $\omega_d$ are expected, the dependence of $\lvert \alpha \rvert$ on $\phi_{d}$ may be considered surprising.
Typically, the drive phase $\phi_{d}$ dictates the phase-space quadrature of the displacement, as opposed to its magnitude -- this is why phase-sensitive measurement is challenging in many QHO systems.
Here, the sensitivity to phase can be understood as the result of interference between the displacement due to the dipole and blue quadrupole tone with the displacement from the dipole and red quadrupole tone (Fig.~\ref{fig:qvsa}(c)).
If the phase sensitivity is undesirable, such as in cases of shot-to-shot dipole phase fluctuations, the electric field can still be faithfully measured by applying only one quadrupole tone.

\begin{figure}
    \centering
    \includegraphics[width = 1.0\textwidth]{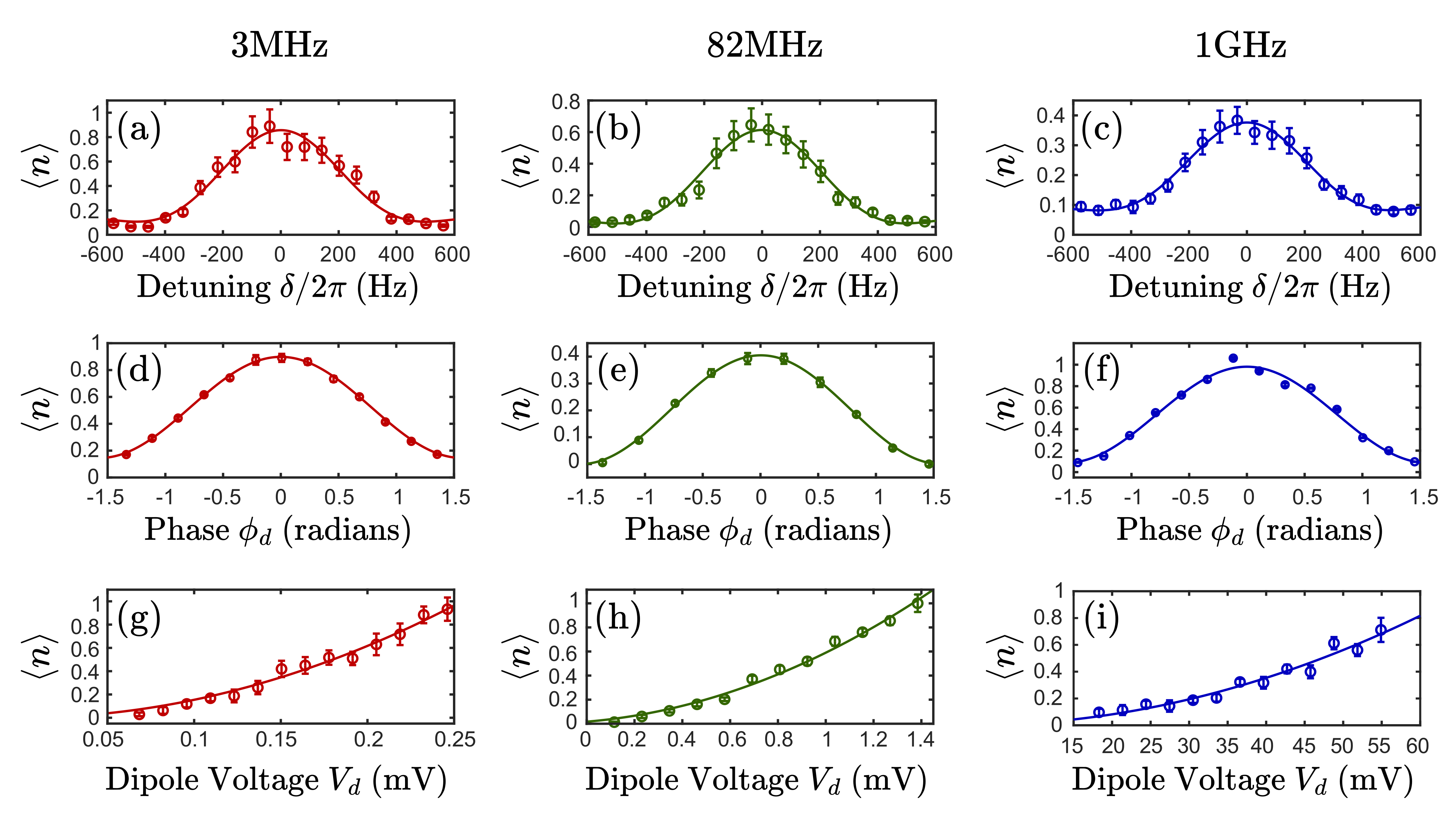}
    \caption{Wideband demonstration of frequency, phase, and amplitude sensing using the QVSA. (a)-(c) Dipole signals are detected by sweeping a pair of quadrupole tones across $\omega_d/{2\pi}$ = 3MHz, 82MHz, and 1GHz, respectively. The lineshape $\langle n\rangle = \lvert \alpha \rvert^{2}$ is obtained by varying the detuning $\delta$. (d)-(f) illustrate the dependence of $\langle n\rangle$ on $\phi_d$ for 3MHz, 82MHz, and 1GHz respectively. The phase dependence is measured with  detuning $\delta=0$. The quadratic dependence of $\langle n\rangle$ on $V_d$ is measured in (g)-(i). The combination of frequency, phase, and amplitude sensing provide the means for vector signal analysis. Solid lines are fit using Eq.~\ref{eq2} to extract voltages experienced by the ion. Error bars represent one standard error. 
    Exact parameters used for each panel are listed in Tab.~\ref{tab:table2} of the SI.
    }
    \label{fig:wbqvsa}
\end{figure}

Using this dependence, we demonstrate wideband vector signal analysis of an unknown dipole tone $(V_d,\omega_d,\phi_d)$ as follows.
First, the QHO is prepared in the ground state $\ket{0}$.
Next, the quadrupole tones are applied at a chosen amplitude $\Omega_q$ and detuning $\delta$ for a time $t$, creating a coherent state $\ket{\alpha}$. 
Finally, the mean phonon number $\langle n\rangle = \lvert \alpha \rvert^{2}$ of the QHO is measured from the relative intensity of the red and blue motional sidebands of the optical qubit transition (see Appendix E of the SI). 
This process is repeated for different detunings $\delta$ to maximize the displacement, yielding the unknown frequency $\omega_d$ (Fig.~\ref{fig:wbqvsa}(a)-(c)).
Once $\omega_d$ is determined, $\phi_d$ is found by repeating the above protocol at $\delta = 0$ while varying the phase of the applied quadrupole tones.
As shown in Fig.~\ref{fig:wbqvsa}(d)-(f), the displacement is maximized when $\phi_d$ is zero.
With $\omega_d$ and $\phi_d$ determined, the measurement of $\langle n\rangle$ can be used to determine $V_d$.
This measurement is shown in Fig.~\ref{fig:wbqvsa}(g)-(i) as the strength of the dipole tone is varied and the quadrupole voltage kept constant.

Two further measurements are reported in Appendix E of the SI.
First, to benchmark the system and showcase its use as a vector network analyzer, we measure the transfer function of a commercial low pass filter (Mini-Circuits: SLP-100+).
Second, as this technique reports the field seen by the QHO, it naturally provides a method for \emph{in situ} calibration of the filter function of a qubit control line if the qubit is coupled to a QHO.
The present experimental apparatus has been designed to demonstrate electric-field gradient gates on molecular ion qubits~\cite{EGGs} and we have therefore used this technique to measure the coupling efficiency of qubit control signals from our frequency synthesizers to the ion.
This technique could also be applied to measure the filter function of electrodes used for shuttling in the quantum charge-coupled device (QCCD) trapped-ion quantum computer~\cite{Moses2023}, as well as for superconducting qubits coupled to a microwave resonator.

\begin{figure}
    \centering
    \includegraphics[width = 1.0\textwidth]{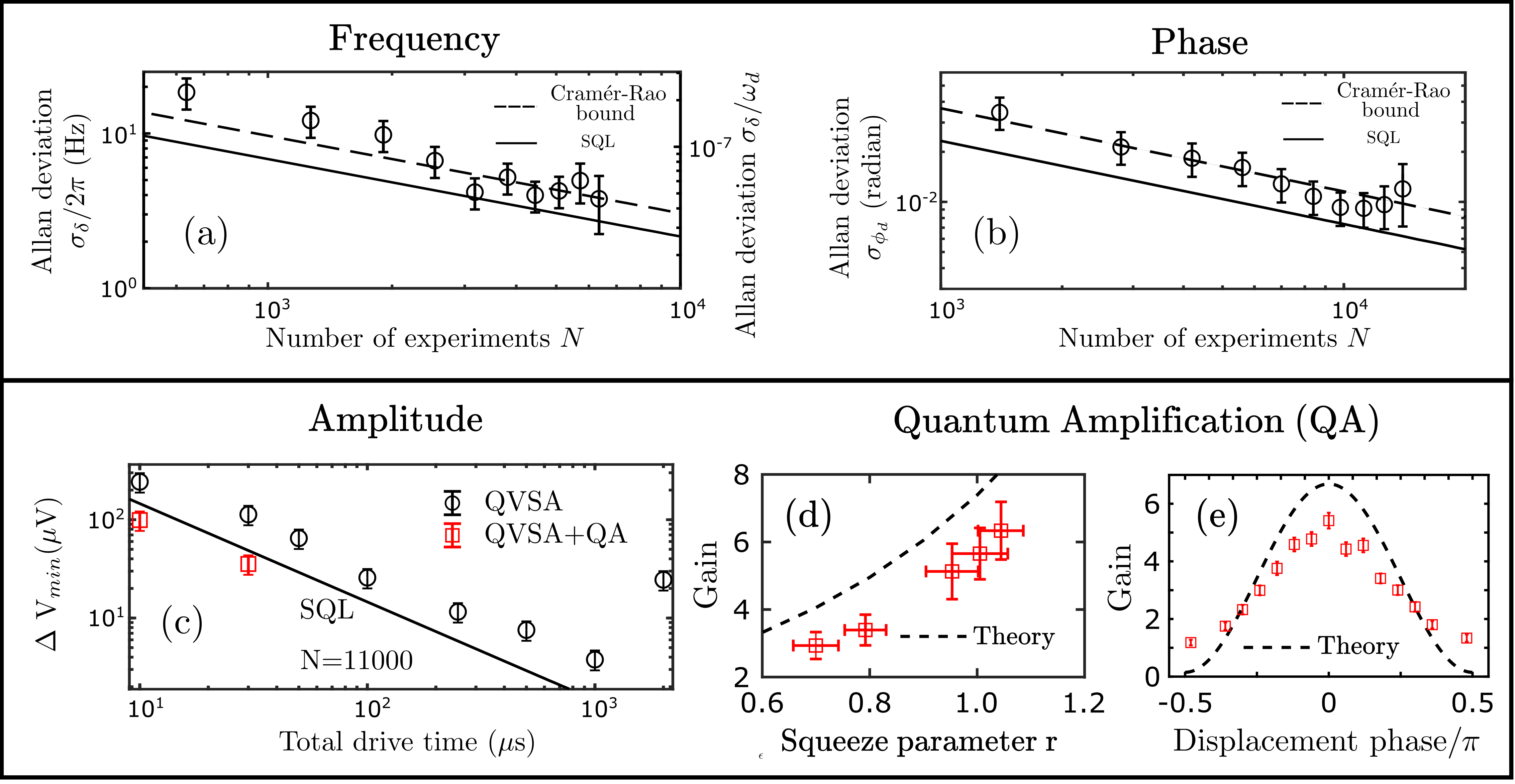}
    \caption{Frequency, phase and amplitude sensitivity of QVSA and integration of quantum amplification (QA). (a-b) Overlapping Allan deviation of frequency and phase of a 82~MHz dipole tone.
    (a) The measured frequency sensitivity is 39(9)~Hz/$\sqrt{\textup{Hz}}$ with a minimum resolution of 3.8(1.6)~Hz after 6300 experiments.
    (b) The measured phase sensitivity is 117(26)~mrad/$\sqrt{\textup{Hz}}$ with a minimum resolution of $9.1(2.0)$~mrad after 9800 experiments.
    (a-b) the probe time is 2~ms and each experiment takes $\sim$17~ms. 
    Solid lines represent the corresponding standard quantum limit (SQL).
    Dashed lines represent the corresponding Cramér-Rao~(CR) bounds, both of which are saturated by our measurements.
    (c-e) Integration of quantum amplification with the QVSA.
    (c) The minimum detectable voltage ($\Delta$V$_{min}$) on the electrodes with and without quantum amplification.
    Each data point represents 11000 experiments, and takes $\sim17$ ms each.
    The measured sensitivity to the voltage coupled onto the electrodes is \SI{51(10)}{\micro V/\sqrt{\textup{Hz}}} with a minimum detectable voltage of \SI{3.8(8)}{\micro V}. 
    The electric field sensitivity is 66(13)~mV$\textrm{m}^{-1}$/$\sqrt{\textrm{Hz}}$ with an electric field resolution of 4.9(1.0)~mV/m.
    Notably, with QA the voltage sensitivity is 3.4(2.0)~dB and 2.8(2.0)~dB below the SQL for \SI{10}{\micro s} and \SI{30}{\micro s}, respectively. 
    (d-e) Quantum amplification through squeezing.
    By applying a squeeze operation $\hat{S}{\left(r\right)}$ before and an anti-squeeze operation $\hat{S}^{\dag}{\left(r\right)}$ after \SI{10}{\micro s} of QVSA operation, the mean phonon number $\langle n \rangle$ is amplified.
    The gain scales with the squeeze parameter $r$ (d) and its phase relative to the QVSA displacement, with $r=0.95$ (e).
    Dashed lines represent the maximum achievable gain from theory.
    Parametric squeezing is applied for \SI{30}{\micro s} at roughly 75~mV.
    (a-e) Error bars represent one standard error. }
    \label{fig:below_sql}
\end{figure}

With these demonstrations complete, it is natural to explore the precision and sensitivity limits of the technique. 
To determine the frequency precision, a rough spectrum, like that shown in Fig.~\ref{fig:wbqvsa}(a)-(c), is first taken to determine the lineshape.
Then, to balance the sensitivity with the intrinsic noise in subsequent experiments, measurements of the phonon number are repeatedly taken at two points in frequency near the half-maxima to determine the center frequency. 
The resulting overlapping Allan deviation is shown in Fig.~\ref{fig:below_sql}(a) alongside the Cramér-Rao~(CR) bound and the SQL, $\Delta\delta\approx2\pi\cdot0.209/(t\langle\alpha\rangle\sqrt{N})$, where $t = 2$~ms is the duration of the QVSA interaction and $N$ is the number of measurements (see Appendix F of the SI). 
The Allan deviation of the parameter estimator saturates the CR bound and is near the SQL with frequency sensitivity 39(9)~Hz/$\sqrt{\textup{Hz}}$ until $N \approx 6300$, achieving a minimum resolution of 3.8(1.6)~Hz, corresponding to a fractional frequency resolution of $4.6(2.0)\times 10^{-8}$.
Similarly, the precision of the phase measurement is determined by a two-point phase measurement at $45\degree$ and $135\degree$ with on-resonant detuning.
The resulting Allan deviation is shown in Fig.~\ref{fig:below_sql}(b) alongside the CR bound and the SQL, $\Delta\phi_d=1/(\langle\alpha\rangle\sqrt{2N})$ (see Appendix F of the SI).  
The minimum phase resolution reaches $\approx 9.1(2.0)$~mrad after $N=9800$ experiments with sensitivity 117(26)~mrad/$\sqrt{\textup{Hz}}$. 
The deviation from the CR bound is due to changes in the secular frequency $\omega$, which depends on the stability of the temperature and humidity of the lab environment and can drift by $\approx2\pi\cdot(200-600)$~Hz over the course of these measurements.

To determine the sensitivity of the voltage measurement, a large $V_q$ is applied to the electrodes to increase the sensitivity to $V_d$ and the resulting phonon number measured.
The process is repeated until the noise floor is reached (typically $N\sim 10^4$).
Fig.~\ref{fig:below_sql}(c) shows the minimum detectable voltage~($\Delta$V$_{min}$) as a function of drive time $t$, with $V_{q} = 8.4(1)$~V~(28~dBm), which is currently limited by the damage threshold of several %RF
electronic components -- with straightforward upgrades this can easily be increased.
The data follows a 1/$\sqrt{N}$ trend until $\approx 1$~ms, where the secular frequency instability becomes significant.
The measured sensitivity to $V_d$ is \SI{51(10)}{\micro V/\sqrt{\textup{Hz}}} with a minimum detectable voltage of \SI{3.8(8)}{\micro V} at $N = 11000$, corresponding to an electric field sensitivity of 66(13)~mV$\textrm{m}^{-1}$/$\sqrt{\textrm{Hz}}$ with an electric field resolution of 4.9(1.0)~mV/m.

Since the QVSA functions by sensing the displacement of the QHO, its voltage (electric field) sensitivity can be increased by leveraging quantum amplification~\cite{Burd2019, Gilmore2021}.
Adapting quantum amplification to the QVSA requires the generation of a squeezed state, $\hat{S}{\left(r\right)}\ket{0}$, prior to application of the QVSA interaction, via e.g. modulation of the trapping potential. 
Assuming a squeeze parameter $r \in \mathbb{R}$, the position uncertainty of the QHO wavefunction is reduced by $e^{-r}$, while the momentum uncertainty is increased by $e^{r}$~\cite{Natarajan1995,Burd2019}.
Next, the system evolves under the QVSA Hamiltonian to produce a displaced squeezed state $\hat{D}{\left(\alpha\right)} \hat{S}{\left(r\right)} \ket{0}$, where for simplicity we assume $\alpha \in \mathbb{R}$.
After this evolution, the system is subject to an anti-squeeze operation to produce the state $\hat{S}^{\dag}{\left(r\right)} \hat{D}{\left(\alpha\right)} \hat{S}{\left( r\right)}\ket{0}$, which has two effects.
First, the anti-squeeze operator returns the displaced squeezed state to a coherent state, allowing measurement of the total displacement at the SQL. 
Second, the anti-squeeze operator amplifies the displacement by $e^r$ along the position axis, yielding a displacement amplitude of $e^{r} \alpha$ to allow sensing of $V_{d}$ below the SQL.

To demonstrate this amplification, we squeeze the oscillator using a parametric trap drive with a maximal coupling strength $r/t = 2\pi\cdot 5.5(2)$~kHz. 
The anti-squeeze operation is achieved by adding a phase shift of $\pi$ to the same drive.
To verify the fidelity of the anti-squeeze operation, we measure the phonon number when performing anti-squeezing directly following the squeeze operation. For a coupling strength of $r/t = 2\pi\cdot 5.5(2)$~kHz, applied for \SI{30}{\micro s}, we recover $\langle n\rangle = 0.03(2)$ which is indistinguishable from the initial state.
Fig.~\ref{fig:below_sql}(d) shows the resulting gain in $\langle n \rangle$ from quantum amplification, measured as a function of $r$.
As described above, maximum amplification occurs when the QVSA displacement is oriented along the squeezing axis in phase space.
The relative phase between the QVSA displacement operator and the squeeze operator can be varied without affecting the phase difference with the dipole tone by simultaneously varying the phase of the blue and red quadrupole tones by the same amount but with opposite sign (see Appendix B of the SI), as shown in Fig.~\ref{fig:below_sql}(e), or by changing the phase of the parametric drive.
As can be seen in Fig.~\ref{fig:below_sql}(d) a maximum gain in $\langle n \rangle$ of 6.3(9) is achieved.
The discrepancy between the achieved and predicted gains (dashed lines) appears to be a result of phase instability between the squeezing and displacement operations.
We characterize the voltage sensitivity in the manner previously described and find that the sensitivity is 10$\mathrm{log}_{10}{\left(\Delta \mathrm{V}_\textup{QA}/\Delta \mathrm{V}_\textup{SQL}\right)}^{2}=$ 3.4~(2.0)~dB and 2.8(2.0)~dB below the SQL for QVSA drive times of \SI{10}{\micro s} and \SI{30}{\micro s}, respectively (see Fig.~\ref{fig:below_sql}(c)).
Operation at longer timescales in our current system is limited by the increased motional decoherence of a squeezed state.

\begin{figure*}
    \centering
    \includegraphics[width = 1.0\textwidth]{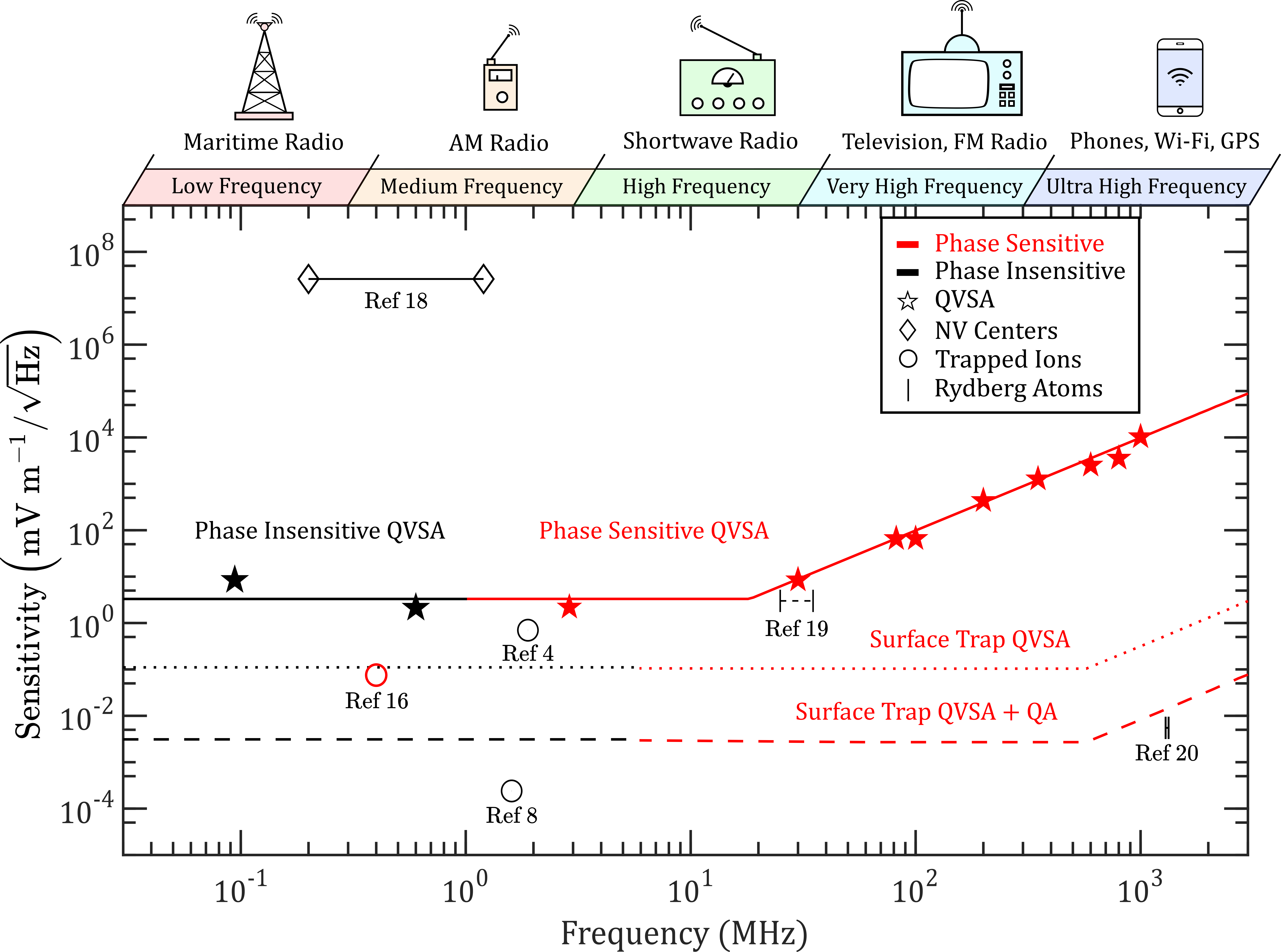}
    \caption{
        Comparisons of electric field sensitivity between different quantum sensing techniques. Phase insensitive methods are shown in black; phase sensitive methods are shown in in red; $\diamond$: NV centers;
         $\circ$: ions; $\vert$: Rydberg atoms; $\star$: our measurement.
         The length of lines corresponds to the frequency range of operation.
        The solid red line represents the estimated QVSA performance in our linear trap, which aligns closely with our measurements (exact measurement results are listed in Tab.~\ref{table1} of the SI).
        Quadrupole tones are applied with a constant Mathieu $q$-parameter of $q_{q} \approx 0.01$ to yield a nearly constant sensitivity from $100 \textrm{ kHz}$ until $\approx 20 \textrm{ MHz}$.
        Above $\approx 20 \textrm{ MHz}$, practical limitations on power prevent the constant Mathieu q-parameter of $q_{q} = 0.01$ from being, causing the minimum achievable sensitivity to increase with frequency.
         The dotted red line indicates the estimated QVSA sensitivity if implemented on a surface trap with an ion-electrode distance of \SI{40}{\micro m}, Be$^+$ as a trapped ion, a secular frequency of $2\pi \cdot 6$ MHz, and a maximum quadrupole power of 39.5~dBm; 
         The dashed red line represents anticipated near-term enhanced QVSA sensitivity in a surface trap with quantum amplification (QA).
         Further details on these improvements can be found in Appendix G of the SI.
     }\label{fig:state_of_art_comp}
\end{figure*}

To demonstrate the wideband capability of the QVSA, we measured the electric field sensitivity at 10 additional frequencies ranging from 100 kHz to 1 GHz, shown as stars in Fig.~\ref{fig:state_of_art_comp}.
 As can be inferred from Eq.~\eqref{eq2}, and further detailed in Eq.~\ref{eq:sensitivity_improvement} of the SI, in the limit of large $\omega_d$ the quantity $q_{q} = \frac{2 e V_{q}}{m r_{0}^{2} \omega_{q}^{2}}$, which interestingly is the Mathieu-q parameter of the \emph{quadrupole} fields, determines both the sensitivity and the bandwidth of the technique. 
However, there are two practical limitations to choosing $q_q$ in our system.
First, we have experimentally determined that in our current system $q_q \leq 0.01$ is required to prevent the quadrupole tones from disturbing the ion trap dynamics. 
This limitation sets the maximum $V_q$, and therefore the sensitivity, for detection at frequencies $\lesssim 20$~MHz. 
At frequencies $\gtrsim 20$~MHz, the required $V_q$ to maintain $q_q = 0.01$ necessitates a power in excess of the damage threshold (28 dBm) of electronic components used in our experiment.
For this reason, $q_q = 0.01$ cannot be maintained above $\approx 20$~MHz and the sensitivity of the technique, in the present implementation, deteriorates with increasing frequency. This is reflected in the solid red line in Fig.~\ref{fig:state_of_art_comp}, where the change in slope at $20~\textrm{MHz}$ is due to this practical inability to maintain $q_q = 0.01$, resulting in an increase in minimum sensitivity with frequency.
This limit is not fundamental.

To contextualize the performance of the QVSA technique, Fig.~\ref{fig:state_of_art_comp} presents a comparison of electric field sensitivity across various quantum sensing techniques.
The QVSA is shown to work over a frequency range $>800\times$ larger than previous techniques and closely matches the performance expected from Eq.~\eqref{eq:sensitivity_improvement} (solid red line) - see Appendix G in the SI for further details.
The phase sensitivity is maintained until $\omega_d \lesssim \omega$, as only the blue quadrupole sideband can be applied in this regime, thereby preventing the interference that generates phase sensitivity.
However, phase sensitivity could be recovered by interfering the displacement from the blue-sideband Raman transition with a resonant displacement at $\omega$. 
Also shown in Fig.~\ref{fig:state_of_art_comp} are the results achieved via other methods.
Ref.~\cite{Qiu2022} reports phase-insensitive electric field measurements in NV centers with a frequency range of roughly 200~kHz to 1~MHz.
Both Ref.~\cite{Wolf2019} and Ref.~\cite{Gilmore2021} are phase-insensitive measurements using either 150 entangled ions in a Penning trap \footnote{This measurement was not scaled by the ion number for comparison to the present single ion results} or a single ion in a Paul trap, respectively, are limited to operating at the oscillator's motional frequency.
Ref.~\cite{affolter2020} demonstrates the ability to perform highly sensitive measurements of the electric field strength, even for fields oscillating at frequencies $\sim$MHz away from the oscillator's motional frequency, with phase sensitivity.
While Ref.~\cite{Meyer2020} predicts that Rydberg atoms can theoretically provide a tunable frequency range from kHz to THz, to our knowledge, the only experimental demonstrations within the MHz to GHz range are Ref.~\cite{Liu2022} and Ref.~\cite{Holloway2022}, which are phase-insensitive measurements demonstrating a frequency range of 10~MHz and 4~MHz, respectively, both shown in Fig.~\ref{fig:state_of_art_comp}.
In comparison, the QVSA is the only method with phase sensitivity, the ability to operate across the low-frequency RF to UHF microwave bands, and demonstrates sensitivities competitive with more specialized techniques.

Several straightforward modifications to the system could significantly improve the performance over that demonstrated here.
These include using a lighter mass ion and smaller electrode-ion distance (e.g. Be$^+$ with a surface trap~\cite{McCormick2019}), as well as higher damage threshold electronics - these form the red dotted line in Fig.~\ref{fig:state_of_art_comp}.
If longer motional coherence times can be achieved, quantum amplification can be leveraged for even greater improvements in sensitivity - this forms the red dashed line in Fig.~\ref{fig:state_of_art_comp}.
Other modifications such as  
the use of other non-classical motional states ~\cite{Hempel2013,Johnson2017,Monroe1996, Wolf2019} and 
a low-noise pre-amplifier could achieve a frequency sensitivity of $\approx 0.3$~Hz/$\sqrt{\textrm{Hz}}$, a phase sensitivity of $\approx 3$~mrad/$\sqrt{\textrm{Hz}}$, and an amplitude sensitivity of $\approx 0.2$~pV/$\sqrt{\textrm{Hz}}$~(corresponding to \SI{2.8}{\micro V~\textrm{m}^{-1}/\sqrt{\textrm{Hz}}}), see Appendix G of the SI for further details.
Finally, we have found that as $V_q$ is increased, the quadrupole tones generate a squeezing interaction \emph{intrinsic} to the QVSA Hamiltonian, which creates an intrinsically phase-squeezed state~(see Appendix C of the SI); this will be the subject of future work where it will be harnessed for \textit{intrinsic} amplification.

In conclusion, we have demonstrated a quantum vector signal analyzer using a single \ca{} ion that is capable of making high sensitivity measurements of the frequency, phase, and amplitude of electric fields in the 100~kHz to 1~GHz range.
This scheme uses Raman transitions between the QHO Fock states to heterodyne a field of interest into the response band of the QHO, where it can be measured.
It can be extended to other QHO architectures with the ability to measure displacements and implement the necessary Raman transitions, such as the Penning trap of Ref.~\cite{Gilmore2021}, where our scheme could be employed to broaden the frequency range in the search for dark matter. 
Beyond trapped ions, our scheme could be implemented in solid-state platforms: by coupling a microwave tone to a superconducting Josephson parametric amplifier (JPA) at frequency $\omega_d$, and inductively coupling a pump tone to the DC superconducting quantum interference device (SQUID) of the JPA~\cite{Yamamoto2008} at frequency $\omega_q$, the full QVSA Hamiltonian could be realized.
This may allow sensitive electric field detection at frequencies $> 1$~GHz and \emph{in situ} calibration of superconducting/ion qubit control lines -- such calibrations are notoriously challenging due to e.g. the stray capacitance of measurement lines and insufficient generality or directness of existing techniques~\cite{Jerger2019,Milne2019, gely2023insitu}.
Further, the QVSA technique could be leveraged to produce an effective beam-splitter-like Hamiltonian to entangle microwave photons with motional phonons, and could therefore be used for frequency transduction (see Appendix H of the SI) or the sensing of electric field noise spectra.
Finally, the higher motional frequencies of trapped electrons~\cite{Matthiesen2021} could extend the detection band to THz signals, while the use of superconducting resonators with cryogenic amplifiers may allow sensitivities significantly below the current state-of-the-art.

\section{Acknowledgement}
We thank S. C. Burd, P. Hamilton, and W. C. Campbell for helpful discussion. 
This work was supported by NSF (PHY-2110421 and OMA-2016245), AFOSR (130427-5114546), and ARO (W911NF-19-1-0297). 

\section{Author contributions}
H.W. and E.R.H. conceived and designed the protocol. 
H.W., G.D.M., and C.Z.C.H. built and maintained the experimental setup and carried out the measurements. E.R.H. supervised all work.
All authors discussed the results and contributed to the manuscript.

\section{Competing interests}
The authors declare no competing interests.

\bibliography{ref}

\appendix
\onecolumngrid
\include{SI}

\end{document}

%% file: SI.tex
\section{Theoretical Analysis}
\subsection{Appendix A: The QVSA Propagator}
Suppose an unknown, oscillatory signal at frequency $\omega_d$ is applied to the trap in a dipole configuration while two signals at frequencies at $\omega_d + \delta \pm \omega$ are applied to the trap in a quadrupole configuration, as shown in Fig.~1(a) of the main text.
We assume $\omega_d \gg \omega\gg \delta$.
The motion of the ion can described by the Hamiltonian:
\begin{align}
    \hat{H}/\hbar =  \omega\hat{a}^\dagger\hat{a} + \Omega_q (\hat{a}+\hat{a}^\dagger)^2
    \cos{\left( \left( \omega_d + \delta \right) t \right) } \cos{ \left( \omega t \right) } + \Omega_d(\hat{a}+\hat{a}^\dagger) \cos{\left(\omega_d t + \phi_d \right)}
\end{align}

where $\Omega_q = eV_q x_o^2/(\hbar r_o^2)$, $\Omega_d = \kappa eV_d x_o/(\hbar  r_o)$, $r_o$ is the ion-electrode distance, $x_o=\sqrt{\frac{\hbar}{2m\omega}}$ is the length scale of the QHO, and $\kappa$ is a geometrical factor. 
In the interaction picture with respect to the harmonic oscillator we have:
\begin{align}
    \hat{H}_{I}/\hbar =&~\Omega_q \left( \hat{a}^2 e^{-2\imath\omega t}+2\left(\hat{a}^\dagger\hat{a} + \frac{1}{2}\right)+ \hat{a}^{\dagger 2} e^{2\imath\omega t} \right) \cos{ \left( \left( \omega_d + \delta \right) t \right) } \cos{ \left( \omega t \right) } \nonumber \\
    &+ \Omega_d \left( \hat{a} e^{-\imath\omega t} +\hat{a}^\dagger e^{\imath \omega t} \right) \cos{\left( \omega_d t+\phi_d \right)}.
    \label{eq:interaction_hamiltonian_full}
\end{align}

To evaluate the propagator, we use the Magnus expansion, i.e. $\hat{U} = e^{\sum_i \hat{A}_i}$.
The first order Magnus term $\hat{A}_{1}$ consists of terms with amplitudes that scale as $\Omega_q/\omega_d$ and $\Omega_d/\omega_d$, which we assume to be small and thus ignore.

The second order Magnus term is given as: 
\begin{align} \label{eq:magnus_2}
    \hat{A}_2 = \frac{1}{2}&\left(-\frac{\imath}{\hbar}\right)^2 \int_0^t\int_0^{t'} \left[ \hat{H}(t'),\hat{H}(t'')\right] dt'' dt'.
\end{align}
Again, we assume that terms proportional to $\Omega_q^{2}/\omega_d^{2}$ and $\Omega_d^{2}/\omega_d^{2}$ are small and lead to trivial light shifts, and thus can be ignored (see Appendices B and C for a more complete treatment). This leaves only the terms that arise from the dipole-quadrupole interaction, which are responsible for the primary dynamics of interest.
\par After applying a RWA to drop terms that oscillate at $2\omega$ and $2\omega_d$ and evaluating the time integrals, the propagator is evaluated as:
\begin{align}
    \hat{U} &= \exp \left[ \imath  \Omega_q\Omega_d \frac{\omega}{\omega_d^2-\omega^2}\frac{\sin{\left(\delta t -\phi_{d}\right)} + \sin{\left(\phi_d\right)}}{\delta}\left(\hat{a} + \hat{a}^\dagger\right) \right] = \hat{D}{\left(\alpha\right)}
    \label{eq:displacement_qvsa_full}
\end{align}
which is the form of a displacement operator with $\alpha \equiv \imath \Omega_q\Omega_d \frac{\omega}{\omega_d^2-\omega^2}\frac{\sin{(\delta t -\phi_d) + \sin{\left(\phi_{d}\right)}}}{\delta}$. In the $\delta t \rightarrow 0$ limit, \eqref{eq:displacement_qvsa_full} reduces to:
\begin{align}
    \alpha = \imath \Omega_{q}\Omega_{d} \frac{\omega}{\omega_{d}^{2}-\omega^{2}} t \cos{\left(\phi_{d}\right)}.
    \label{eq:displacement_qvsa_small_dt}
\end{align}

\subsection{Appendix B: The QVSA Propagator with Phase Delay}
Defining the RSB (BSB) phase as $\phi_r$ ($\phi_b$), the Hamiltonian of the ion motion is:
\begin{align}
    \hat{H}/\hbar  =&~ \omega\hat{a}^\dagger\hat{a} + \Omega_d \left( \hat{a}+\hat{a}^\dagger \right) \cos{\left(\omega_d t+\phi_d \right)} \nonumber \\
    & + \frac{\Omega_q}{2} \left(\hat{a}+\hat{a}^\dagger\right)^2 \left( \cos{\left(\left(\omega_d + \delta + \omega\right)t+\phi_{b}\right)} + \cos{\left(\left(\omega_d + \delta - \omega \right)t+\phi_r\right)}\right).
\end{align}
As in Appendix A, the resulting propagator is a displacement operator with:
\begin{align}
    \alpha = \imath\Omega_q\Omega_d \frac{\omega}{\delta(\omega_d^2-\omega^2)} e^{\imath \phi_-}(\sin(\delta t-\phi_d+\phi_+)+\sin(\phi_d-\phi_+)) \label{eq:appendix_a*_displacement_1}
\end{align}
where $\phi_{\pm}=\frac{\phi_b \pm \phi_r}{2}$ (cf. \eqref{eq:displacement_qvsa_full}). Assuming $\delta t \ll 1$, this becomes:
\begin{align}
    \alpha = \imath\Omega_q\Omega_d \frac{\omega t}{\omega_d^2-\omega^2} \frac{e^{\imath (\phi_d-\phi_r)}+e^{-\imath (\phi_d-\phi_b)}}{2}.
    \label{eq:appendix_a*_displacement_2}
\end{align}
The first (second) exponential in \eqref{eq:appendix_a*_displacement_2} is the phase term corresponding to the red (blue) sideband tone.
Setting $\phi_r=\phi_b=0$ (following the main text) motivates the choice of phase for Fig.~1(b).

\begin{figure}[h!]
    \centering
    \includegraphics[width = 0.9\textwidth]{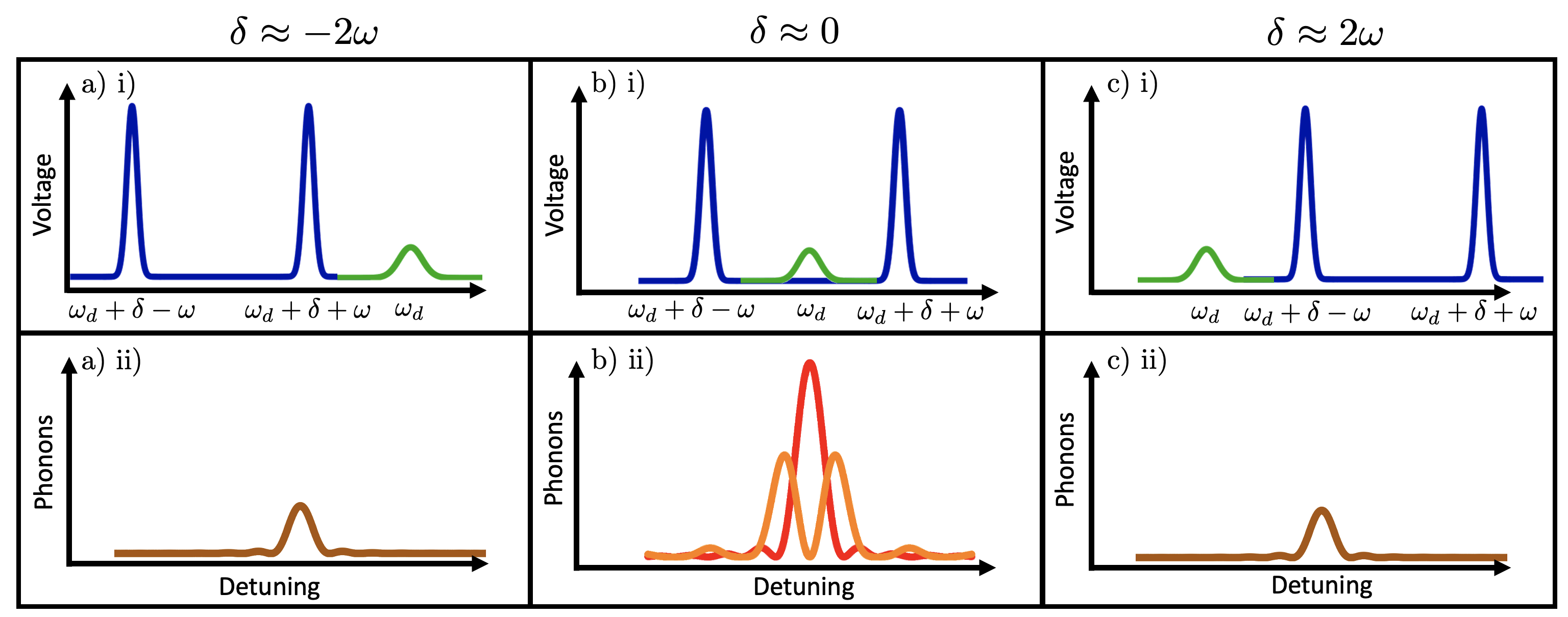}
    \caption{
    Illustration of spectrum sweep procedure.
    (a-c, i) Two quadrupole tones (blue) are swept in frequency across a weaker dipole tone (green).
    (a-c, ii) The resulting mean phonon number $\langle n \rangle$ produced from the corresponding interaction in (i).
    (a, i-ii) The interaction is primarily determined by that between the blue quadrupole tone at $\omega_{d} + \delta + \omega$ and the dipole tone at $\omega_{d}$. This interaction is phase-independent.
    (b, i-ii) Both quadrupole tones at $\omega_{d} + \delta \pm \omega$ interact with the dipole tone at $\omega_{d}$. The resulting interaction is phase-dependent, and can produce a resonance (red) when the phase relation satisfies $\phi_d-\phi_+=0$, where $\phi_{+} = \frac{\phi_{b} + \phi_{r}}{2}$, or an antiresonance (orange) when $\phi_d-\phi_+=\frac{\pi}{2}$.
    (c, i-ii) Conversely to (a), the interaction is primarily determined by that between the red quadrupole tone at $\omega_{d} + \delta - \omega$ and the dipole tone at $\omega_{d}$, and is phase-independent.
    }
    \label{fig:sweep}
\end{figure}

\begin{figure}[h!]
    \centering
    \includegraphics[width = 1\textwidth]{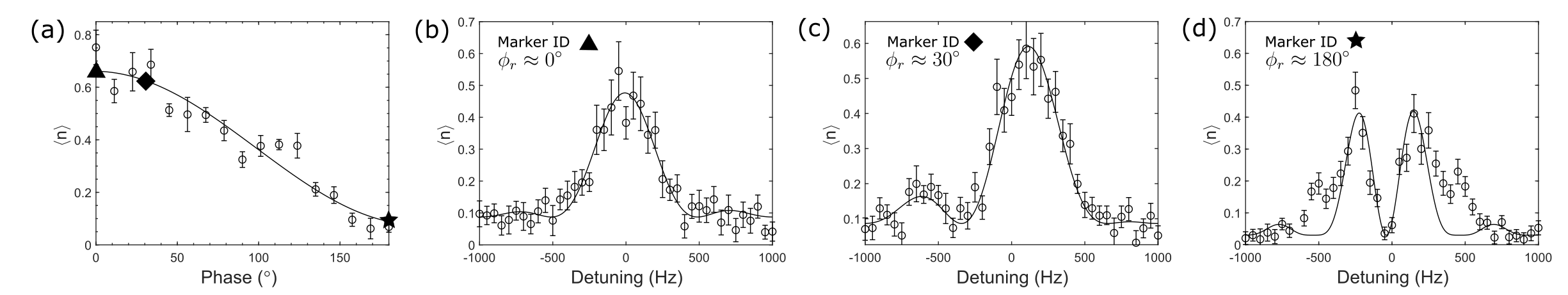}
    \caption{Phase dependence of the spectrum sweep procedure.
    The phase relation between the QVSA tones fundamentally changes the lineshape of the resonance.
    By sweeping just $\phi_r$, the phase dependence is reduced to $\langle n \rangle \propto \cos^2{(\frac{\phi_r}{2})}$. 
    (a) shows a phase sweep at the resonance (i.e. $\delta = 0$), with three characteristic phases marked by a triangle (b), a diamond (c), and a star (d). Voltage parameters vary between marked points.
    (d) illustrates the production of an antiresonance, which occurs when the red sideband is completely out of phase with the blue sideband (see Fig. 1(b) in the main text for a graphical representation). The amplitude discrepancy between (a) and (b-d) is due to a difference in the applied QVSA voltages. Error bars represent one standard error.
    The solid lines are fitted lines using \eqref{eq:appendix_a*_displacement_1}. 
    \label{fig:phase_depend_antiresonance}
    }
\end{figure}

As the frequencies of the quadrupole tones are scanned across the dipole frequency, three peaks are observed (Fig.~\ref{fig:sweep}). 
Two peaks arise at $\delta \approx \pm 2 \omega$ from a phase-independent interaction between the dipole tone and each quadrupole tone individually, while a central peak at $\delta \approx 0$ arises from the phase-dependent interaction between the dipole and both quadrupole tones (Fig.~\ref{fig:phase_depend_antiresonance}).
While scalar spectrum analysis is therefore possible with only a single sideband tone, \textit{vector} signal analysis requires the presence of all three tones to yield a phase dependence, and has the added benefit of a stronger interaction.

\newpage

\subsection{Appendix C: Intrinsic Squeezing}
Consider the effect of the two quadrupole tones in isolation, i.e. $V_{d} = 0$. For simplicity, we choose $\phi_{r} = \phi_{b} = 0$. The interaction Hamiltonian of the ion motion is (cf. \eqref{eq:interaction_hamiltonian_full}):
\begin{align}
    \hat{H}_{I} / \hbar = \Omega_{q} \left( \hat{a}^{2} e^{-2i \omega t} + 2 \left( \hat{a}^{\dagger} \hat{a} + \frac{1}{2} \right)+ \hat{a}^{\dagger 2} e^{2i \omega t} \right) \cos{\left( \left( \omega_d+\delta \right) t \right)} \cos{\left( \omega t \right)}.
\end{align}
As in Appendix A, we consider only the second order Magnus term and apply a RWA to find the propagator:
\begin{align}
    \hat{U} \approx & \exp \left[ \frac{i \Omega_{q}^{2}t}{2}\left(\frac{\omega}{\omega_d^{2}-\omega^{2}}\right) (\hat{a}^{2} + \hat{a}^{\dagger 2}) \right] = \hat{S}{\left( z \right)}
    \label{eq:quadrupole_squeezing_propagator}
\end{align}
which is the form of a squeezing operator with $z \equiv -i\Omega_{q}^{2} \left(\frac{\omega}{\omega_d^{2}-\omega^{2}}\right) t$.

We now consider the interaction of all three tones in full. Combining \eqref{eq:displacement_qvsa_full} and \eqref{eq:quadrupole_squeezing_propagator} gives the second order Magnus term:
\begin{align}
    \hat{A}_{2} = \left(\alpha \hat{a}^{\dag}-\alpha^{*}\hat{a}\right) + \frac{1}{2} \left(z^{*}\hat{a}^{2} - z\hat{a}^{\dag 2} \right)
\end{align}
where $\alpha$ is defined in \eqref{eq:displacement_qvsa_small_dt}.

Using the Zassenhaus relation, the propagator of the interaction can be expressed: 
\begin{align}
    \hat{U} &= \hat{S}{\left(z\right)} \hat{D}{\left(\sum_{j=1}^\infty \frac{\imath^j(-1)^{j+1}\lvert \alpha\rvert\lvert z\rvert^{j-1}}{j!} \right)}
    = \hat{S}{\left(z\right)} \hat{D}{\left(\alpha'\right)} \label{eq:intrinsic_squeezing_propagator} \\
    &\textrm{where}~\alpha' \equiv \Big|\frac{\alpha}{z}\Big|\left(\imath\sinh{\lvert z \rvert} - \cosh{\lvert z \rvert} + 1\right).
    \label{eq:intrinsic_squeezing_displacement}
\end{align}
The complete QVSA interaction thus consists of two parts: a displacement operator $\hat{D}{\left(\alpha'\right)}$ and a residual squeeze operator $\hat{S}{\left(z\right)}$.

The resulting displacement $\alpha'$ is larger than that calculated in \eqref{eq:displacement_qvsa_small_dt} due to the intrinsic squeezing generated by the quadrupole tones \eqref{eq:quadrupole_squeezing_propagator}, an effect we refer to as \textit{intrinsic amplification}.
For small quadrupole voltages, i.e. as $\Omega_{q} / \omega_{q} \rightarrow 0$, the squeezing parameter becomes negligible, i.e. $z \rightarrow 0$, and thus $\alpha' \rightarrow \alpha$, reproducing \eqref{eq:displacement_qvsa_small_dt} and \eqref{eq:displacement_qvsa_full}.

\begin{figure}[h!]
    \centering
    \includegraphics[scale = 0.7]{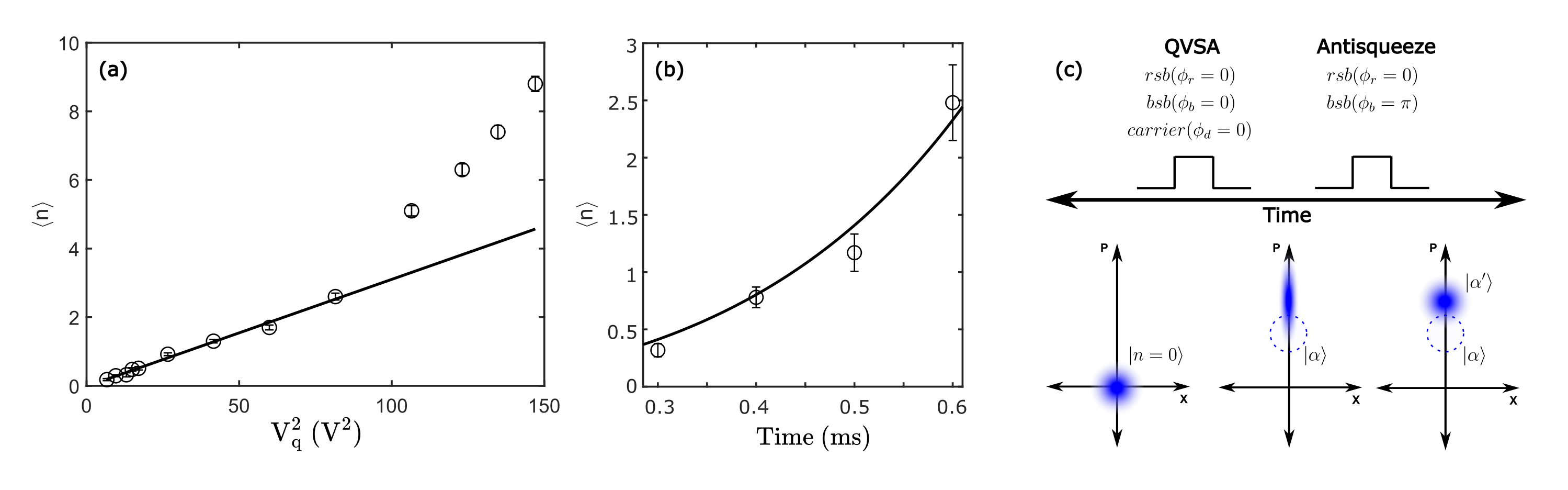}
    \caption{Onset of intrinsic amplification in the high $V_q$ regime.
     (a) Heating from the QVSA interaction as a function of the quadrupole voltage $V_{q}$.
     % Intrinsic amplification at high probing voltages.
     The solid line is a linear fit to the data in the low $V_q$ regime.
     The deviation of the data from the solid line in the high $V_{q}$ regime arises from the intrinsic squeezing of the QVSA interaction, and can be leveraged to improve the measurement sensitivity.
     (b) Characterization of the intrinsic squeezing strength. The solid line is a fitted line using \eqref{eq:quadrupole_squeezing_propagator} and yields a coupling strength $\lvert z \rvert/t \approx 2\pi\cdot300$ Hz.
     Error bars represent one standard error.
     (c) Pulse sequence for intrinsic amplification and associated Wigner functions.
     The dashed lines on the bottom figures depict a pure displacement to $\ket{\alpha}$ as would be expected from the solid line in (a).
     The QVSA pulse is comprised of a displacement operation followed by a squeeze operation, as illustrated by the colored ellipse.
     To convert from a squeezed state back to a coherent state $\ket{\alpha'}$ we antisqueeze by shifting the blue sideband phase by $\pi$.
     }
    \label{fig:intrinsic_amp}
\end{figure}

The residual squeezing results in an undesirable increase in the measurement background, which becomes particularly pronounced for longer measurement times (Fig. \ref{fig:intrinsic_amp}(b)). % the creation of a displaced-squeezed state leads to 
However, the residual squeezing can be negated by applying an antisqueezing operation $\hat{S}^{\dag}{\left( z\right)}$ following the QVSA interaction (\textit{intrinsic antisqueezing}). In practice, this can be straightforwardly implemented by disabling the dipole tone and phase-shifting a single quadrupole tone by $\pi$.

\newpage
\section{Experimental Details}

\subsection{Appendix D: Experimental Procedures}
\subsubsection{QVSA Sequence}
\begin{figure}[h!]
    \centering
    \includegraphics[scale = 0.75]{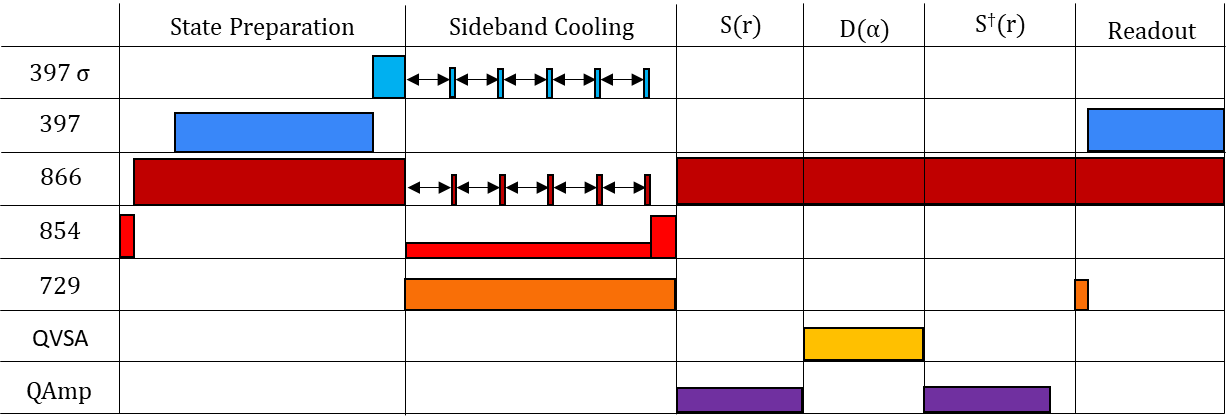}
    \caption{Pulse sequence for the QVSA with Quantum Amplification.
    The ion is Doppler cooled and optically pumped into the $\ket{^{2}S_{\frac{1}{2}},m_J=-\frac{1}{2}}$ state.
    Continuous sideband cooling subsequently prepares the ion near the ground motional state $\ket{n=0}$.
    Quantum Amplification is achieved by preparing the ion in a squeezed state before applying the QVSA pulse, which amplifies the displacement and minimizes uncertainty along the squeezing axis.
    The squeezing is then reversed via an anti-squeezing pulse, returning the ion to a coherent state.
    }
    \label{fig:pulse}
\end{figure}

Each QVSA experiment is comprised of four steps. A sample pulse sequence including quantum amplification is shown in Fig. \ref{fig:pulse}.

First, the ion is prepared in the motional ground state via sideband cooling (Fig.~\ref{fig:coherent}(b)).
Second, the QVSA interaction is applied to induce a displacement $\alpha$.
Third, the motional state is projected onto the $\{S,D\}$ spin manifold via excitation on the first red/blue motional sideband for some given time $t$. This procedure can be expressed as a unitary operation $\hat{U}_{r/b}$. For sideband ratio measurements, we choose $t = t_{\pi} = \pi / \Omega_{01}$, where $\Omega_{01}$ is the Rabi frequency for the $\ket{S,n=0}$ to $\ket{D,n=1}$ transition.
The spin state can in turn be read out via state-dependent fluorescence, which are described by the projection operators $\{ \hat{\Pi}_{\mu} \}$ where $\mu \in \{ 0 \text{ (dark)}, 1 \text{ (bright)} \}$.
The final D-state probability can thus be expressed as $P_{r/b} = P_{r/b}\left(\mu|\alpha\right) = $ Tr$\{\hat{\Pi}_{\mu}\hat{U}_{r/b}^{\dag} \hat{\rho}(\alpha) \hat{U}_{r/b}\}$.
Fourth, the ratio of the red and blue sideband D-state populations~($R=P_r$/$P_b$) is converted to a displacement and a mean phonon number at different detunings of the quadrupole tones (relative to the dipole tone), assuming a coherent state (Fig.~\ref{fig:coherent}(a, c-f)).
The resulting profile is fitted with a sinc$^{2}$ function to extract the displacement (Fig.~\ref{fig:coherent}(f)).
However, for Allan deviations, we instead measure only a single point at the secular frequency resonance where $\delta_{q} = \omega$.

\subsubsection{Signal Generation}
Signals for the QVSA interaction are generated using the Sinara ``Phaser" Arbitrary Waveform Generator (AWG).
The AWG has two coherent output channels which are used to separately address a pair of quadrupolar trap electrodes (Fig. \ref{fig:schematic}).
Dipole tones (i.e. $\hat{x}$ terms) and quadrupole tones (i.e. $\hat{x}^{2}$ terms) can be simply generated from the same AWG by setting an inter-channel phase delay of $\pi$ and $0$, respectively.

In practice, the QVSA signals are applied to excite motion \textit{perpendicular} to the QVSA electrodes (i.e. along the y mode), which is made possible due to imperfections in trap geometry, though the effective dipole voltage is reduced by an experimentally determined geometric factor $\kappa = 1/42$. 
This is desirable since the x mode experiences a higher motional heating rate as a result of the electronics connected to the QVSA electrodes.

\subsubsection{Calibration Procedures}
Phase delays and latencies between the signals applied to the QVSA electrodes change the multipole moments of the field and can thus reduce the QVSA interaction strength and the motional coherence time. However, these can be simply calibrated by sweeping the inter-electrode phase delay at the RF source to maximize the resulting displacement.

Similarly, the QVSA signals can be aligned in phase space (Fig. \ref{fig:qvsa}(b)) by sweeping the phase of a single quadrupole tone (e.g. $\phi_{r}$, the phase of the red quadrupole tone) to maximize the resulting displacement. From \eqref{eq:appendix_a*_displacement_1}, maximal displacement occurs when $\phi_{d} = \phi_{+} = \frac{\phi_{r} + \phi_{b}}{2}$, where $\phi_{r}$ and $\phi_{b}$ are the red and blue quadrupole phases, respectively.

Calibration of the quadrupole amplitudes is achieved by applying a dipole tone together with a single quadrupole tone and equalizing the resulting displacement (Fig.~\ref{fig:sweep}(a, c)). The quadrupole voltages can in turn be determined absolutely by applying the two quadrupole tones at $\omega_{d} \pm \omega$ without the dipole tone to generate a motional squeezed state (see Appendix C), from which the squeezing coupling strength (and thus the voltage) can be extracted (Fig.~\ref{fig:intrinsic_amp}(b)).

\subsubsection{Sample Applications}
\begin{figure}[h!]
    \centering
    \includegraphics[width = 0.95\textwidth]{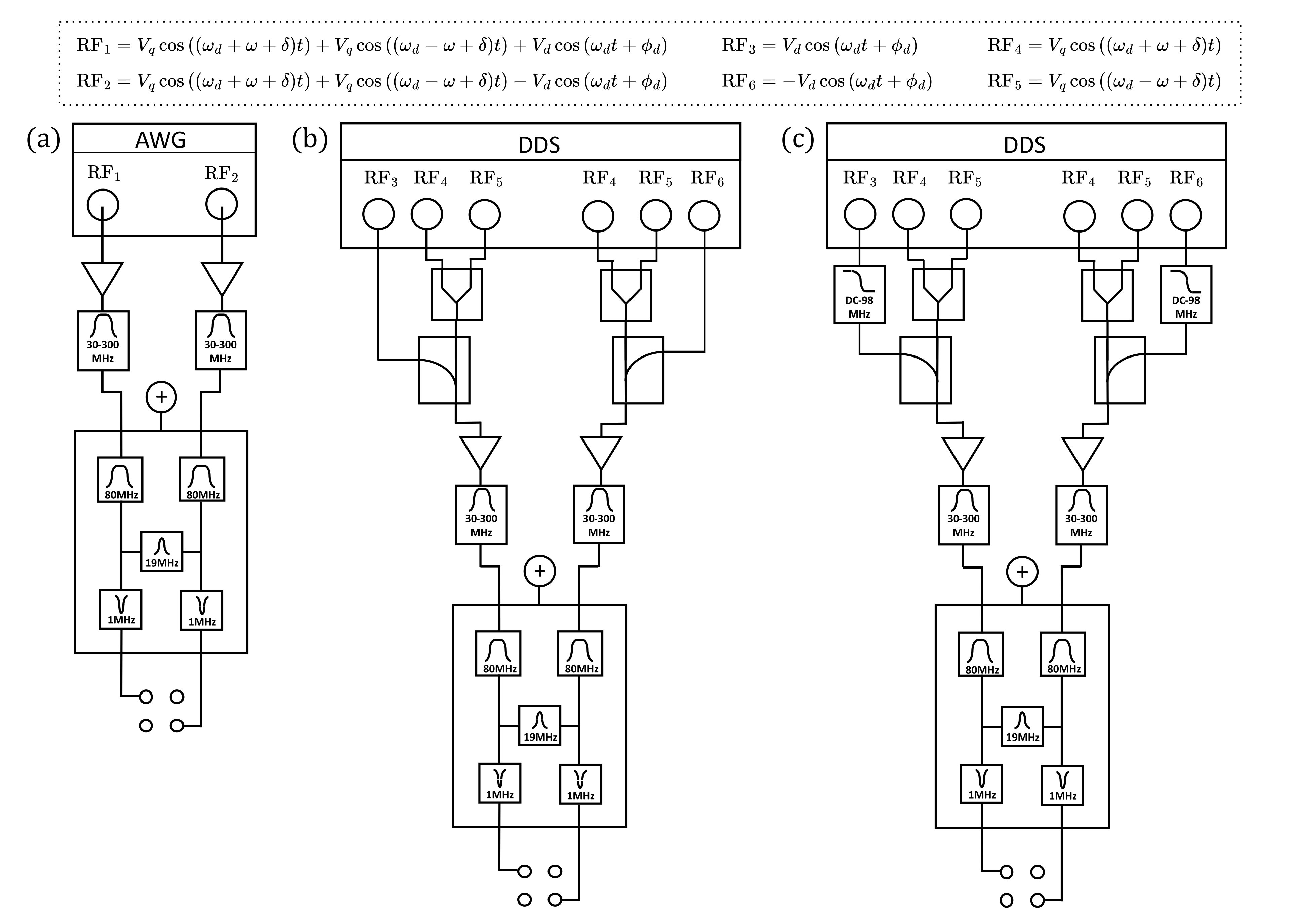}
    \caption{
    Example RF schematic of the QVSA signal generation apparatus.
    (a-c) Signals are amplified (Mini-Circuits ZHL-20W-13SW+) before filtering and DC biasing (to break the radial mode degeneracy). A narrow 19 MHz band-pass filter between the QVSA electrodes shorts any pickup from the 19 MHz trap electrodes, which improves trap stability. Different combinations of filters can be used for different frequency ranges. 
    (a) Dipole and quadrupole tones are generated with two channels of an Arbitrary Waveform Generator (Sinara ``Phaser") by setting an inter-channel phase delay of $\pi$ and $0$, respectively.
    (b-c) QVSA signals are generated using two banks of synchronously clocked Direct Digital Synthesizers (DDSs) (Sinara AD9910 Urukul). This configuration allows QVSA function as a vector network analyzer (VNA) as the dipole tones can be separately passed through a Device Under Test.
    }
    \label{fig:schematic}
\end{figure}

\begin{figure}
    \centering
    \includegraphics[width = 0.6\textwidth]{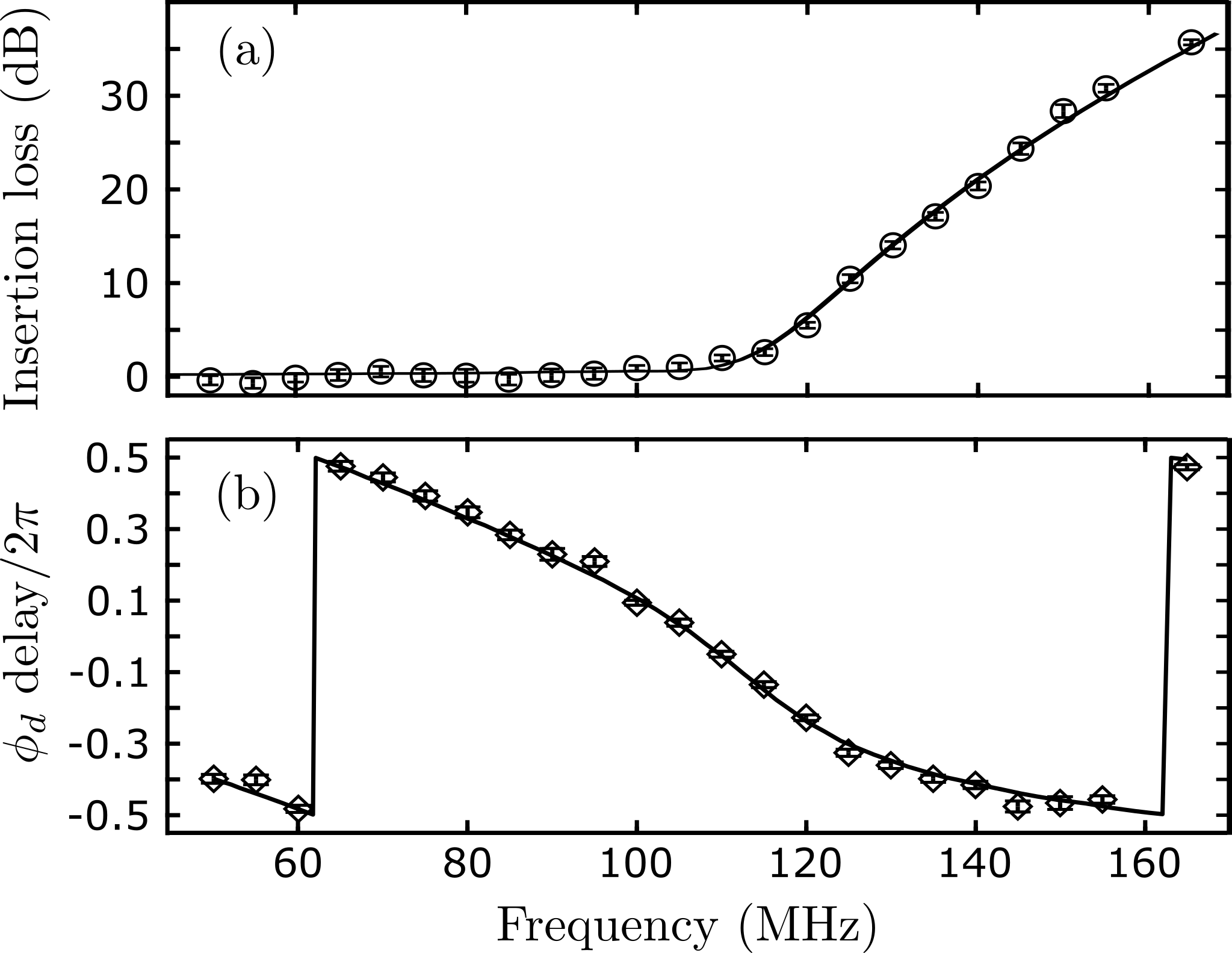}
    \caption{
        Transfer function of a commercial low pass filter.
        (a-b) $\circ$ and $\diamond$ are measured through the QVSA.
        Solid lines are measured via a commercial vector network analyzer.
        Each point is composed of $N \approx 13500$ experiments.
        Error bars represent one standard error.
        No re-scaling or offset is applied to the data. 
    }
    \label{fig:bode_plot}
\end{figure}

The relevance of the QVSA technique is demonstrated through two examples.
We demonstrate the functionality of the QVSA technique for vector network analysis by characterizing the transfer function of a commercial low-pass filter (Mini-Circuits SLP-100+). 
For this measurement, $V_d$ and $\phi_d$ are measured as a function of frequency with and without the filter inserted.
The RF configuration used for these measurements is shown in Fig.~\ref{fig:schematic}(b-c).
For normalization, the transfer function of the dipole tone is measured without the Device Under Test (DUT) (i.e. the low-pass filters) in place (Fig.~\ref{fig:schematic}(b)).
At each frequency, the RSB phase $\phi_{r}$ is scanned, and the resulting displacement is fitted to Eq.~\eqref{eq2} in the main text to extract the voltage ($V_n$) and phase ($\phi_n$) of the measured signal, similarly as shown in Fig.~\ref{fig:wbqvsa}.
Next, DUTs are connected in series with the relevant dipole tone DDS channels(RF3 and RF6) respectively as shown in (Fig.~\ref{fig:schematic}~(c)), which alter the amplitude and phase of the dipole field experienced by the ion.
A similar measurement is repeated to extract the new voltage ($V_f$) and phase ($\phi_f$) sensing by the ion.
Finally, the phase delay from low pass filters can be expressed as ($\phi_f$-$\phi_n$) and attenuation can be written as $20 \textrm{log}_{10}(V_f/V_n)$.
The data is shown in Fig.~\ref{fig:bode_plot} alongside the measurement from a standard network analyzer (Agilent 8714ES), which shows excellent agreement. 

\begin{figure}[h!]
    \centering
    \includegraphics[width = 0.8\textwidth]{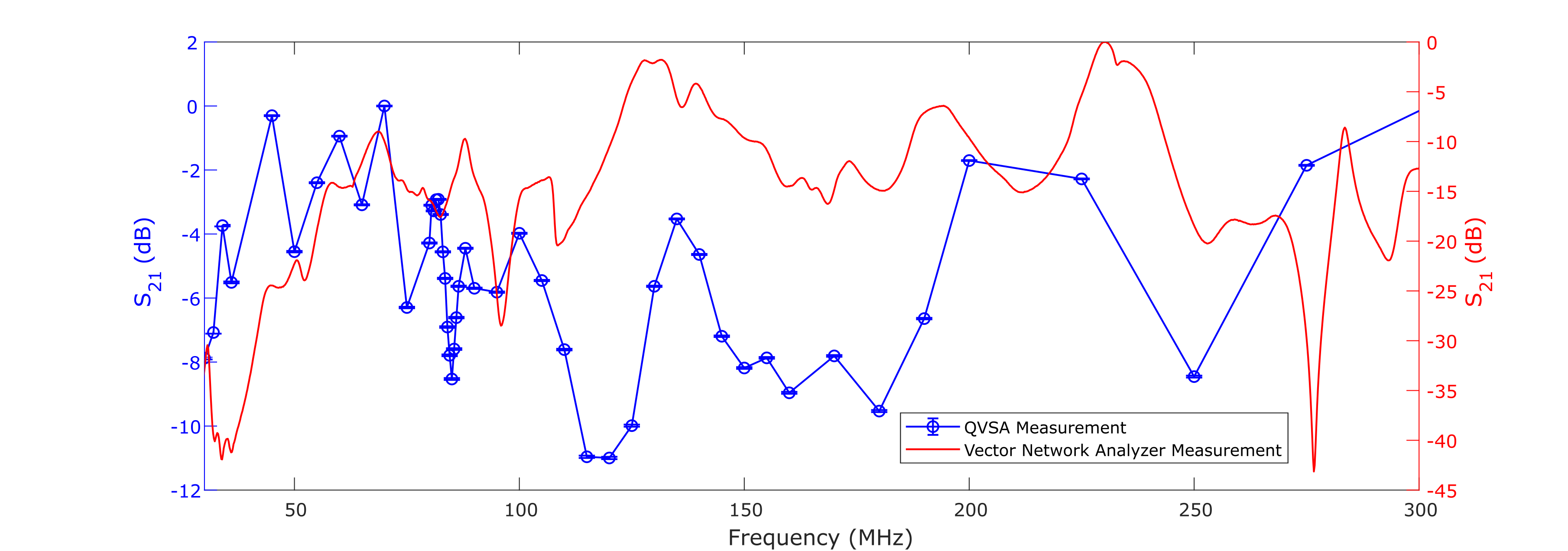}
    \caption{Insertion loss of the ion trap system measured using the QVSA and a commercial vector network analyzer (VNA).
    Blue (red) data shows the insertion loss measured by the QVSA (VNA).
    Error bars for the QVSA measurements represent one standard error and are too small to be seen.
    }
    \label{fig:bode}
\end{figure}

The same procedure was also used to calibrate the wideband transfer function of a qubit control line (Fig.~\ref{fig:bode}).
The measured coupling efficiency is compared against that from a traditional measurement using a capacitor divider connected to trapping RF electrodes and a vector network analyzer over the frequency range accessible with our DDS (20~MHz - 300~MHz).
As shown in Fig.~\ref{fig:bode}, the traditional measurement fails to accurately report the filter function. 
Changes to the measurement conditions required to interface the network analyzer, e.g. impedance changes at measurement ports, likely account for this discrepancy, serving as an archetypal example of the difficulty of accurately calibrating qubit signals.

\subsection{Appendix E: Motional Detection}
To date, the most sensitive measurement technique for displacement sensing is the overlap technique \cite{Wolf2019}, which employs Fock states to measure the expected state overlap $\lvert \langle n \lvert \hat{D}\left( \alpha \right) \lvert n \rangle \lvert^{2}$.
However, due to the technical complexity of the overlap technique, we instead employ two more straightforward and well-established detection techniques: sideband ratio measurement and blue sideband (BSB) Rabi oscillations.

\begin{figure}[h!]
    \centering
    \includegraphics[width = 0.9\textwidth]{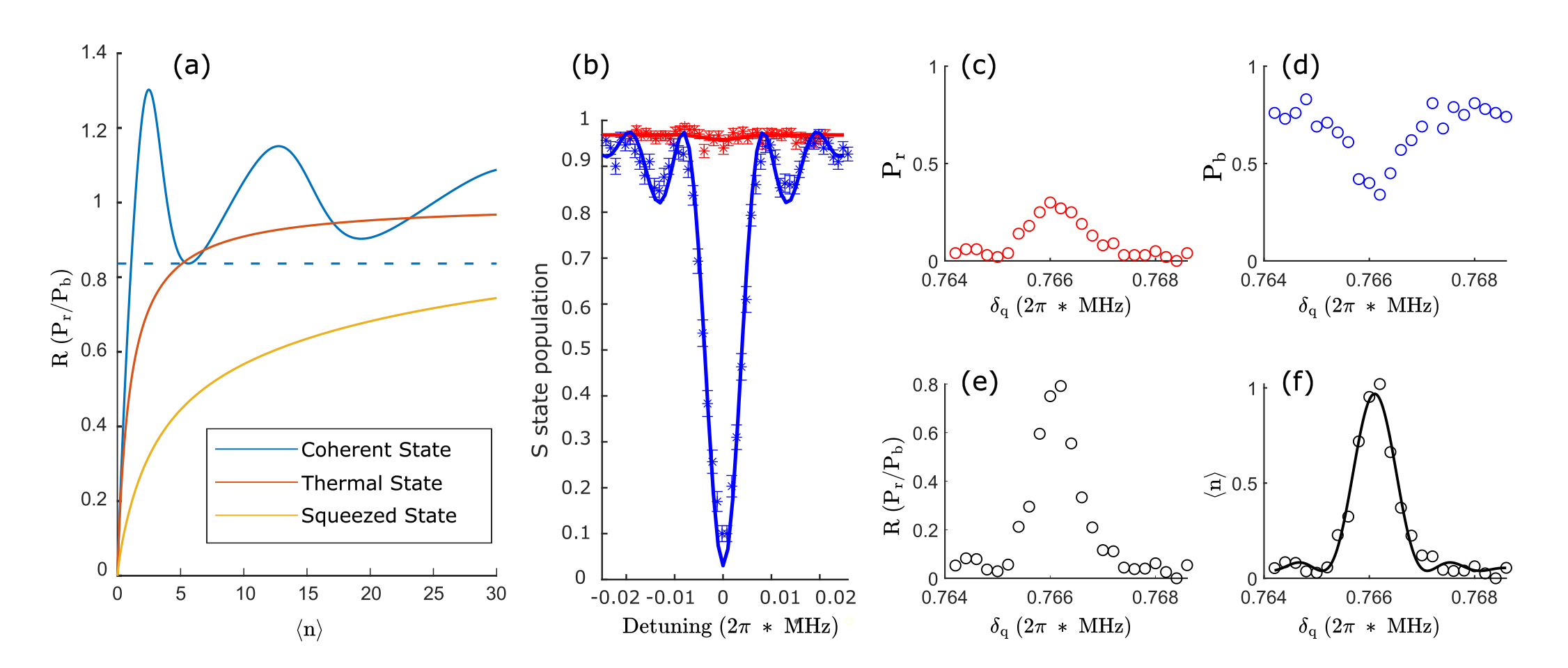}
    \caption{
    Sideband ratio measurement of $\langle n \rangle$.
    (a) Conversion of the sideband ratio $R$ to a mean phonon number $\langle n \rangle$ for different motional states. The sidebands are interrogated for time $t_{\pi} = \pi / \Omega_{01}$, where $\Omega_{01}$ is the Rabi frequency for the $\ket{S,n=0}$ to $\ket{D,n=1}$ transition. To ensure injectivity as $R \rightarrow \langle n \rangle$, we enforce $R < 0.829$ for coherent states, depicted by the dashed line.
    (b) A typical red/blue sideband spectrum after ground state cooling. Assuming a thermal state, the spectra shows cooling to $\langle n \rangle = 0.01 \pm 0.01$.
    Error bars represent one standard error.
    (c-f) Typical sideband spectra after the QVSA interaction is applied and the quadrupole detunings $\lvert\omega_q-\omega_d\rvert$ scanned.
    (c) and (d) are the red and blue sideband spectra, respectively.
    (e) shows the resulting sideband ratio. The QVSA strength has been chosen such that $R < 0.829$. The ratio is converted to mean phonon number $\langle n \rangle$ in (f).
    }
    \label{fig:coherent}
\end{figure}

The sideband ratio measurement is preferred when the motional state is known to be coherent due to its high Fisher Information (Fig.~\ref{fig:coherent}).
For other or unknown motional states, BSB Rabi oscillations are preferred due to their generality and ability to distinguish between types of motional states.

\subsubsection{Sideband Ratio Measurement}
\begin{figure}[h!]
    \centering
    \includegraphics[width = 1\textwidth]{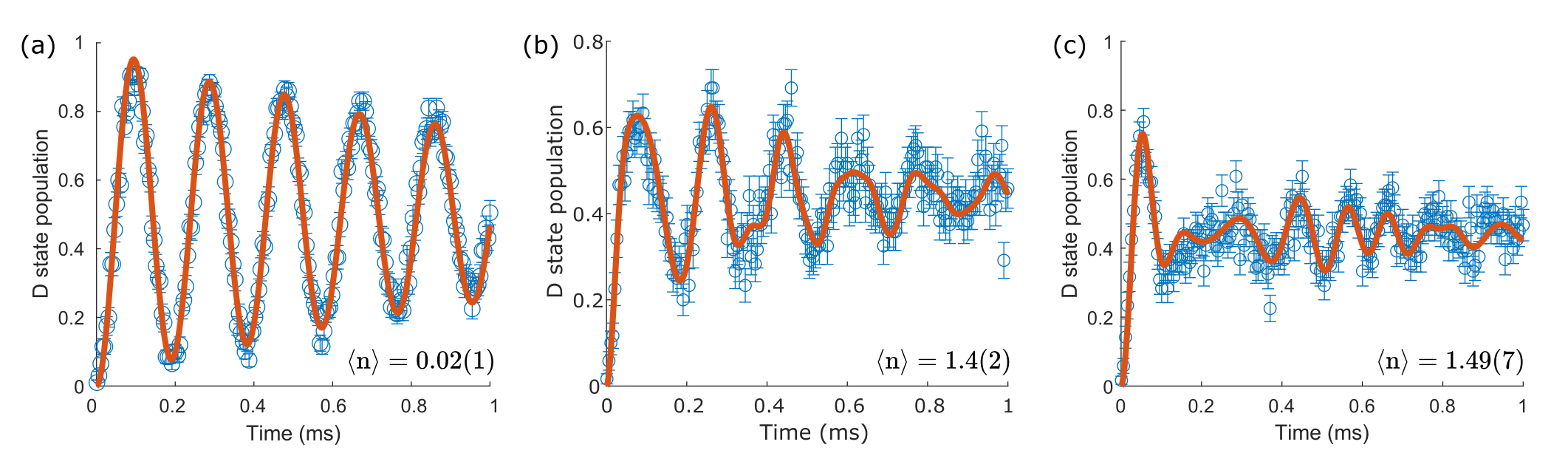}
    \caption{Blue sideband (BSB) Rabi oscillation for different motional states.
    The coherence time $t_{dec}$ is roughly $1400$~$\mu$s, and the Rabi frequency ($\Omega_{00}$) on resonance with the qubit transition  is $2\pi \cdot 110$~kHz. 
    (a) BSB Rabi oscillations after sideband cooling to the motional ground state and fitted using a thermal state.
    (b) BSB Rabi oscillations for a squeezed state, which is prepared through parametric modulation of the trap potential at twice the secular frequency.
    (c) BSB Rabi oscillations for a coherent state produced using the QVSA interaction.
    Error bars represent one standard error.
    }
    \label{fig:bsb_oscill}
\end{figure}

 The sideband ratio technique compares the excitation probabilities of the first red and blue sidebands. This technique is technically simple and offers immunity to systematic sources of error (e.g. imperfect state preparation, qubit frequency drifts, 729nm power fluctuations).
 However, the sideband ratio technique becomes non-injective when $\langle n \rangle > 1$, and is statistically inefficient when applied to squeezed states (Fig. \ref{fig:bsb_oscill}).

\subsubsection{Blue Sideband (BSB) Rabi Oscillations}
Motional detection via BSB Rabi oscillations is accomplished by fitting the oscillations to their theoretical form to extract the relevant state parameter and the Fock state populations \cite{Meekhof1996,Leibfried2003,Lo2015}.
In contrast to the sideband ratio technique, this technique offers dynamic range and generality at the cost of lower Fisher Information and concomitant sensitivity: it can be used for measurements of $\langle n \rangle > 1$ and can be applied to an arbitrary motional state.
We use this method specifically for quantum amplification data, which often results in $\langle n \rangle > 1$ and involves characterization of squeezed states.
The BSB coherence time is $\approx 1400~\mu$s, which is limited by imperfect control, e.g. servo bumps \cite{Akerman2015} and laser power fluctuations of the qubit addressing beam, and systematic drifts, e.g. secular frequency drifts and qubit frequency fluctuations.

\subsection{Appendix F: Statistical Analysis}
\subsubsection{Fisher Information and the Cramér-Rao Bound}
In parameter estimation, the Cramér-Rao Bound (CRB) sets a lower bound on the achievable variance of an estimator $\Delta \alpha$ in terms of its Fisher Information $F$, i.e. $\Delta \alpha \geq \Delta \alpha_{CR}= 1/\sqrt{F}$.

Classically, the Fisher Information is calculated:
\begin{align}
    F{\left( \alpha \right)} = \displaystyle \sum_{x}{\frac{1}{P\left(x|\alpha\right)} \left(\frac{\partial P\left(x|\alpha\right)}{\partial \alpha}\right)^{2}}
    \label{eq:classical_FI}
\end{align}
which is additive for repetitive independent measurements.
The maximal $F$ over the estimator space is called the quantum Fisher information ($\mathfrak{F}$), which in turn leads to the quantum Cramér-Rao bound (QCRB).

\subsubsection{Saturating the Cramér-Rao Bound}
Here, we show that the CRB of our sideband ratio measurement can be saturated by adjusting the relative number of red and blue sideband measurements to constitute $R$.

We define the inversion estimator~\cite{xu2019} $\hat{\alpha} = f^{-1}{\left(R\right)}$ to evaluate the uncertainty of the displacement $\alpha$ given that $R = P_{r}/P_{b} = f{\left(\alpha\right)}$.
Assuming that the uncertainties of the red and blue sideband probabilities $\Delta P_{r/b}$ are dominated by quantum projection noise (which follows a binomial distribution and is thus $\Delta P_{r/b} = \sqrt{p_{r/b} \left(1-p_{r/b}\right) /N}$), propagation of uncertainty gives:

\begin{align}
    \Delta \hat{\alpha} = \sqrt{\frac{1}{F_r N_r} \left(\pdv{\alpha}{R}\right)^{2} \left(\pdv{R}{P_r}\right)^{2} \left(\pdv{P_r}{\alpha}\right)^{2} + \frac{1}{F_b N_b} \left(\pdv{\alpha}{R}\right)^{2} \left(\pdv{R}{P_b}\right)^{2} \left(\pdv{P_b}{\alpha}\right)^{2} } \sqrt{N_r+N_b} \label{eq:uncertainty_inversion_estimator}
\end{align}
where $P_r$ ($P_b$) is the D-state population after a red (blue) sideband pulse, $N_r$ ($N_b$) is the number of red (blue) sideband measurements, and $F_r$ ($F_b$) is the Fisher Information per experiment for a red (blue) sideband measurement.

To account for imperfections in the experimental procedure, we use an empirical formula $P^{real}_{r/b} = A\,P_{r/b}{\left(\alpha\right)} + B$, where $A \approx 0.94$ and accounts for decoherence processes, e.g. laser power fluctuations, while $B \approx 0.03$ and accounts for background noise, which for our experiment is dominated by the servo bump of the qubit addressing laser.

\begin{figure}[h!]
    \centering
    \includegraphics[width = 0.75\textwidth]{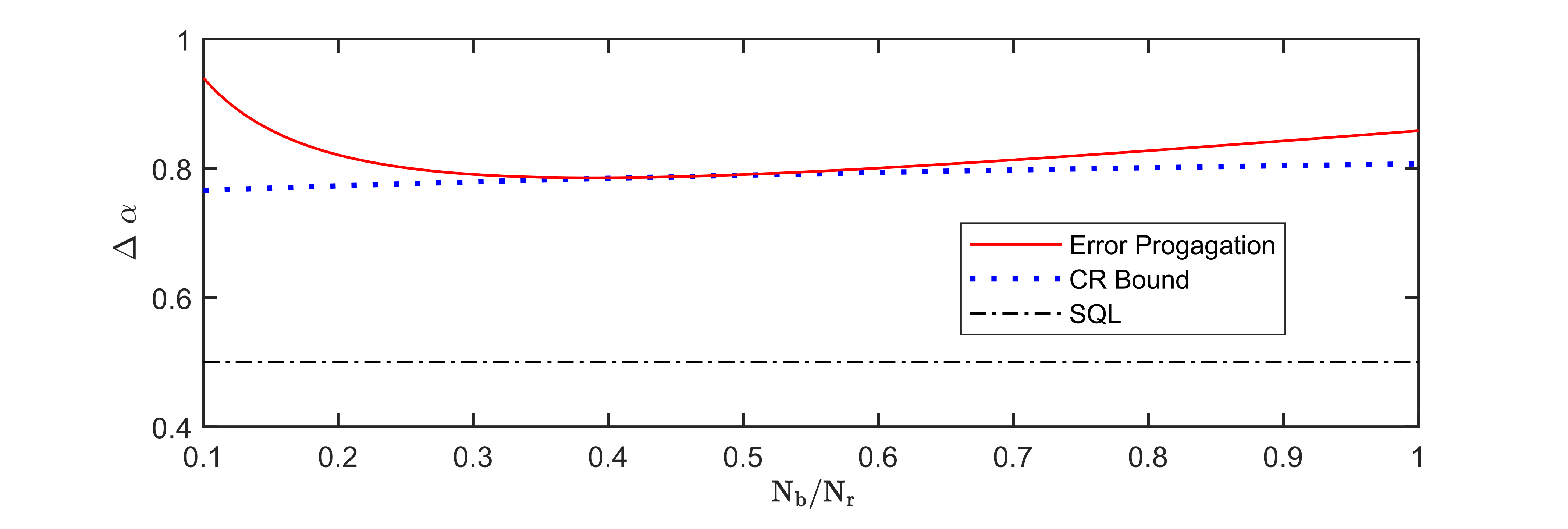}
    \caption{Uncertainty in the measured displacement $\Delta \alpha$ as a function of the relative number of red and blue sideband measurements used for the phase measurement in Fig. \ref{fig:below_sql}(b) in the main text.
    The Cramér-Rao bound (CRB) is saturated when $N_{b}/N_{r} \approx 0.4$. Experimental imperfections are included in the estimation.
    }
    \label{fig:delta_alpha}
\end{figure}

For comparison, the CRB per experiment calculated from \eqref{eq:classical_FI} is:
\begin{align}
    \Delta \alpha_{CRB} = \sqrt{\frac{N_r+N_b}{N_r F_r+N_b F_b}}.
    \label{eq:uncertainty_cr_bound}
\end{align}

Fig.~\ref{fig:delta_alpha} shows both the CRB and $\Delta \hat{\alpha}$ alongside the SQL as a function of $N_b/N_r$ for $\alpha \approx 0.68$. The estimator performance is maximized for $N_b/N_r \approx 0.4$, where it also saturates the CRB.

\subsubsection{The Standard Quantum Limit (SQL)}
The standard quantum limit (SQL) of a measurement is the minimum achievable uncertainty when using classical states.
Here, we derive the standard quantum limit (SQL) for amplitude, frequency and phase sensing as depicted in Fig.~\ref{fig:below_sql} in the main text.

For displacement sensing, the SQL is $\Delta \alpha_{\textrm{SQL}} = \frac{1}{2}$ \cite{Wolf2019, Gilmore2021}.
The SQL for the mean phonon number $\langle n \rangle$ is then:
\begin{align}
    \Delta n_{\textrm{SQL}} &= \frac{\Delta \alpha_{\textrm{SQL}}}{\pdv{\alpha}{n}} = \sqrt{n}.
    \label{eq:sql_n_displacement}
\end{align}

\subsubsection{The Standard Quantum Limit (SQL) - Frequency Sensing}
For frequency sensing with Rabi spectroscopy, the mean phonon number $\langle n \rangle$ expressed in terms of the dipole tone detuning $\delta$ (using Eq.~\ref{eq2} in the main text) is:
\begin{align}
    \langle n \rangle={\lvert \Tilde{a} \rvert}^2{\frac{\sin^{2}{\left(\delta t\right)}}{\left(\delta t\right)^2}}
    \label{eq:n_dipole_freq_relation}
\end{align}
where $\lvert\Tilde{a}\rvert$ is the displacement amplitude and $t$ is the QVSA interaction time. 
Using \eqref{eq:sql_n_displacement}, the SQL becomes $\Delta n = \lvert\Tilde{a}\rvert \sin(\delta t)/(\delta t \sqrt{N})$, where $N$ is the number of experiments. The frequency uncertainty of the dipole tone $\Delta \delta$ at the SQL is therefore:
\begin{align}
    \Delta\delta = \frac{\Delta n}{\pdv{n}{\delta}}
    = \frac{\left(\delta t \right)^2}{\delta t^2 \cos{\left(\delta t\right)} - t \sin{\left(\delta t\right)}} \frac{1}{2 \lvert\Tilde{a}\rvert \sqrt{N}}.
    \label{eq:sql_f_SI}
\end{align}
Unlike in Ramsey spectroscopy, where the uncertainty is independent of detuning~\cite{Itano1993}, Rabi spectroscopy measurements require judicious selection of the readout points. For the Allan deviation in Fig.~\ref{fig:below_sql}(a), we use two points near the half-maxima, which balances the sensitivity with the signal-to-noise ratio~(SNR).
Under these conditions, $\sin{\left(\delta t \right)}/\left(\delta t\right) \approx 1/\sqrt{2}$. The SQL for frequency sensing is thus:
\begin{align}
    \Delta\delta \approx 2\pi \frac{0.209}{\lvert\Tilde{a}\rvert t \sqrt{N}}.
    \label{eq:sql_f_SI_adev}
\end{align}

\subsubsection{The Standard Quantum Limit (SQL) - Phase Sensing}
Following \eqref{eq:n_dipole_freq_relation}, the mean phonon number $\langle n \rangle$ expressed in terms of the dipole tone phase $\phi_{d}$ is:
\begin{align}
    \langle n\rangle = \lvert\Tilde{a}\rvert^{2} \cdot \sin^{2}{\left(\theta+\phi_d\right)},
    \label{eq:n_dipole_phase_relation}
\end{align}
where $\theta$ is the variation of the initial phase. At the SQL, this becomes $\Delta n = \lvert\Tilde{a}\rvert \cdot \sin{\left(\theta+\phi_d\right)} / \sqrt{N}$, which allows us to calculate the phase uncertainty at the SQL:
\begin{align}
    \Delta\phi_{d} = \frac{\Delta n}{\pdv{n}{\phi_d}} = \frac{1}{\cos{\left(\theta+\phi_d\right)}} \frac{1}{2\lvert\Tilde{a}\rvert\sqrt{N}}.
    \label{eq:sql_phi_SI}
\end{align}
Similarly, for the Allan deviation in Fig. \ref{fig:below_sql}(b), two points are measured at the half-maxima (i.e. $\cos{\left(\theta+\phi_{d}\right)}=1/\sqrt{2}$). The SQL for phase sensing is thus:
\begin{align}
    \Delta\phi_{d} = \frac{1}{\lvert\Tilde{a}\rvert\sqrt{2N}}.
    \label{eq:sql_phi_SI_adev}
\end{align}

\subsubsection{The Standard Quantum Limit (SQL) - Amplitude Sensing}
The mean phonon number $\langle n \rangle$ expressed in terms of the dipole tone amplitude $V_d$ is:

\begin{align}
    \langle n \rangle = A V_{d}^{2} t^{2},
\end{align}
where $A = 1.1 \times 10^{13} \  \textrm{V}^{-2} \textrm{s}^{-2}$ is a prefactor (measured using $V_{q} = 8.4 \textrm{V}$ at 82~MHz ). The SQL is thus:

\begin{align}
    \Delta V_{d} = \frac{\Delta n}{\pdv{n}{V_{d}}} = \frac{1}{2 t \sqrt{A N}},
    \label{eq:sql_amplitude}
\end{align}
where $N$ is the number of experiments.

The relationship between voltage ($V_d$) and electric field ($E_d$) is given by $E_d=0.715V_d/r_0$, as determined using COMSOL Multiphysics.

\subsection{Appendix G: Improving Sensitivity}

Generally, the sensitivity to an unknown parameter $\theta$ can be improved by either reducing the uncertainty of the measured quantity (e.g. $\Delta n$), which can be achieved via the use of non-classical states, or by increasing the magnitude of the parameter response (e.g. $\pdv{n}{\theta}$), which can be achieved by adjusting experimental parameters.

\subsubsection{Use of Non-Classical States}
For single QHOs, Fock states, squeezed states, and cat states are the most commonly used non-classical states to probe signals with uncertainty below the SQL. 

The Fisher information for a Fock state $\ket{n}$ is $F^{\textrm{Fock}}_{\alpha} = 8n+4$ \cite{Wolfthesis}. 
Thus, application of the QVSA on a Fock state $\ket{n=12}$ could improve the sensitivity by a factor of $5\times$.

The maximum $F$ for a squeezed state $\ket{r}$ is $F^{\textrm{squeezed}}_{\alpha} = 4 e^{2\lvert r \rvert}$ \cite{Wolfthesis}.
Our maximum achievable squeezing parameter $r$ is limited by our control electronics, improvements to which could yield a gain in sensitivity of $9\times$ \cite{Burd2019}.

The maximum $F$ for a Schrödinger cat state is $F^{\textrm{Cat}}_{\alpha} = 16 \left( \lvert \alpha \rvert^{2} + \frac{1}{4} \right)$\cite{Wolfthesis}. 
Use of a cat state with $\alpha = 3$ could thus improve the sensitivity by roughly $6\times$ \cite{Hempel2013}.

\subsubsection{Experimental Modifications}

To understand the fundamental limits of the QVSA frequency range, Eq.~\ref{eq2} can be made frequency ($\omega_{d}$) independent by instead expressing it in terms of the Mathieu $q$-parameter, as shown below:

\begin{align}   
    \alpha  = \beta \frac{ q_{q} V_{d}}{r_{0} \sqrt{m \omega}} \frac{\sin{\left(\delta t - \phi_{d} \right)} + \sin{\left(\phi_{d}\right)}}{\delta},
    \label{eq:sensitivity_improvement}
\end{align}

where $\beta$ is a constant prefactor and $q_{q} = \{\frac{2 e V_{q}}{m r_{0}^{2} \omega_{q}^{2}} \textrm{ for }  \omega_q \approx \omega_d \gg \omega;\frac{2 e V_{q}}{m r_{0}^{2} (\omega_{d}^{2}-\omega^{2})} \textrm{ for } \omega_d \approx \omega \}$ is the Mathieu $q$-parameter of the quadrupole tones.
This independence of the displacement from $\omega_d$ indicates a potentially infinite bandwidth.

However, at larger frequencies, greater power is required to maintain the same $q_{q}$. On our experiment, we were unable to maintain a fixed $q_{q}$ above $\approx 20 \textrm{ MHz}$ due to damage thresholds of electronic components -- this underlies the change in slope of the expected sensitivity (solid red line) in Fig.~\ref{fig:state_of_art_comp}.
Additionally, empirical evidence on our experiment suggests that trapping dynamics are disturbed once the $q$-parameter of the applied quadrupole tones reaches $10 \%$ of the trap's $q$-parameter -- this effectively limits our implementation to $q_{q} < 0.01$. It should be noted that this constraint is specific to our system -- a higher $q_q$ may be feasible on other systems.

Consequently, there are several straightforward, readily available improvements, all using existing technology, that could greatly enhance QVSA operation. 
The features and benefits of these are outlined below:
\begin{enumerate}
    \item Use of a surface trap \cite{McCormick2019}: a decrease in $r_{0}$ from \SI{550}{\micro m} in our current trap to \SI{40}{\micro m}.
    \item Trapping of a lighter species: e.g. \be, which is lighter than \ca by $\approx 4$.
    \item Stabilization of the secular frequency to $> 10~\textrm{Hz}$: an increase in interaction time from our current $1~\textrm{ms}$ to $> 10~\textrm{ms}$ and greater quantum amplification (which is limited by motional decoherence).
    Currently, motional coherence times up to 55~ms have been achieved\cite{Matsos2024}.
    \item Operation at higher $q_{q}$: the use of a surface trap, with its lower $r_{0}$ would increase $q_{q}$ even with lower input powers. For an $r_{0}$ of \SI{40}{\micro m}, a $200~\textrm{mV}$ field at $82~\textrm{MHz}$ would result in $q_{q} = 0.01$, compared to the $q_{q} = 0.0005$ achieved in our trap with $8.4 \textrm{ V}$.
    Increasing the power damage thresholds of system electronics or the use of a resonator to minimize power requirements would allow higher powers to be applied to achieve large $q_q$ even at high frequencies.
    \item Increase in the secular frequency from $2 \pi\cdot 0.8~\textrm{MHz}$ to $2\pi\cdot6 ~ \textrm{MHz}$: reduction in sideband cooling time from $13~\textrm{ms}$ for \ca to 
    \SI{230}{\micro s} for \be\cite{McCormick2019}, and thus the duration of each experimental trial reduces from $(16+t)~\textrm{ms}$ to $(1+t)~\textrm{ms}$, of which $t$ is QVSA duration.
    This reduction in duration of each experimental trial exceeds the reduction in sensitivity due to the higher secular frequency.
    \item Amplification of the dipole tone: A simple pre-amplifier, e.g. Mini-Circuits ZHL-20W-13SW+ would increase the amplitude sensitivity by $30~\textrm{dB}$.
\end{enumerate}

\subsubsection{Improved Amplitude Sensitivity}
Here, we provide detailed information regarding the estimation of electric field sensitivity enhancement, as illustrated in Fig.~\ref{fig:state_of_art_comp}.

According to Eq.~\ref{eq:sensitivity_improvement}, reducing the $m$ from 40~AMU to 9~AMU, increasing the $q_q$ from 0.01 to 0.03, significantly shortening the time devoted to preparation from 16~ms to 1~ms and increasing the QVSA interaction time from 1~ms to 10~ms collectively enhance the sensitivity by a factor: $\sqrt{\frac{40\textrm{AMU}}{9\textrm{AMU}}} \frac{0.03}{0.01}\frac{10~\textrm{ms}}{1~\textrm{ms}} \sqrt{\frac{0.8\textrm{MHz}}{6\textrm{MHz}}} \sqrt{\frac{17~\textrm{ms}}{11~\textrm{ms}}}$=$30\times$, which leading to an electric field sensitivity 0.1~mVm$^{-1}$/$\sqrt{\textrm{Hz}}$.
An applicable power of 9~W can extend this sensitivity up to 600~MHz, shown in red dotted line in Fig.\ref{fig:state_of_art_comp}.

Assuming increasing the QVSA interaction time further from 10~ms to 55 ms due to enhanced secular frequency stability and expecting at least another factor 2$\times$ gain from quantum amplification (QA) beyond Ref. \cite{Burd2019} leading to 18$\times$.
Then, the sensitivity can be further improved by an additional factor  18$\frac{55~\textrm{ms}}{10~\textrm{ms}}\sqrt{\frac{11~\textrm{ms}}{56~\textrm{ms}}}$=42$\times$.

Following the preceding discussions, the total gain provides an amplitude sensitivity as low as $0.2~\textrm{pV} /\sqrt{\textrm{Hz}}$, yielding in turn an electric field sensitivity of $\approx \SI{2.8}{\micro V~\textrm{m}^{-1}/\sqrt{\textrm{Hz}}}$, which is shown as red dotted line in Fig.\ref{fig:state_of_art_comp}.
This is better than state of art field sensing measurement done with Rydberg atoms \cite{Jing2020}.
We also note that though Ref.\cite{Gilmore2021} is below our sensitivity, it employs 150 trapped ions instead of the one use here. 
To be conservative, we chose not scale this point, nor the points for Rydberg atoms, by the particle number. 
Using more particles, as could be envisioned in a
multiplexed surface trap, would improve our sensitivity even further by $\sqrt{N}$.

\subsubsection{Improved Frequency and Phase Sensitivity}
Unlike amplitude sensitivity, gains in the frequency and phase sensitivities are harder to realize due to their independence from experimental parameters, as is evident from Eq.~\ref{eq:sql_f_SI_adev} and Eq.~\ref{eq:sql_phi_SI_adev} (assuming $\alpha \leq 1$).

With regards to frequency, a five-fold increase in the motional coherence time from $2~\textrm{ms}$ to $10~\textrm{ms}$ would yield a concomitant five-fold improvement in the sensitivity, per \eqref{eq:sql_f_SI_adev}. Implementation of a Ramsey spectroscopy detection scheme, as opposed to the current Rabi spectroscopy detection scheme, would further improve the sensitivity by $\sim 2.5$, as shown in \cite{Wolf2019}. Improvements to our squeezed state generation could further amplify our measurement by a factor of 9.
Together, these modifications could improve the sensitivity by an overall factor of roughly $110\times$ to reach $0.3~\textrm{Hz}/\sqrt{\textrm{Hz}}$.

By contrast, since the phase sensitivity is independent of the QVSA interaction time (see Eq.~\ref{eq:sql_phi_SI_adev}), the only avenue for improvement is through the reduction of the duration of each experimental trial. This time could be reduced from $17~\textrm{ms}$ to $700~\mu\textrm{s}$ for a sensitivity gain of 5. Again, together with the use of a squeezed state, a total gain of roughly $45\times$ in sensitivity could be realized to reach $3~\textrm{mrad}/\sqrt{\textrm{Hz}}$.

\subsection{Appendix H: Field Sensing at the Trap RF Frequency and Parametric Resonances}
In trapped ion experiments, it is known that an electric field with angular frequency $\Omega_{\mathrm{RF}} \pm \omega$, where $\omega$ is the secular frequency of the ion, can heat the ion and reduce sensitivity.
To mitigate this, a high quality factor RF helical resonator ($\Omega_{\mathrm{RF}} = 2\pi \cdot$ 19~MHz, $Q = 120$ and bandwidth: 160~kHz) is employed in our experiment to filter noise at $\Omega_{\mathrm{rf}} \pm \omega$. 
This same filtering leads to a very narrow RF drive and means that good performance of the QVSA outside of the bandwidth of RF resonator can be expected.
If the signal to be measured falls within the bandwidth of the RF resonator, the noise at same frequency can couple onto the trap electrodes via the resonator and increase the background noise.
However, since the field distribution of such noise follows quadrupole configuration, the background noise can be mitigated by moving ion onto the field null - this procedure is well-established and known as micromotion compensation~\cite{Berkeland1998}.
Our QVSA can similarly be used for micromotion compensation and can potentially achieve state-of art resolutions~(3~mV m$^{-1}$/$\sqrt{\textup{Hz}}$), as shown in Fig.~\ref{fig:state_of_art_comp} of the main text.

Beyond that, it is also well known that parametric resonances of the form $\frac{\Omega_{\mathrm{rf}}}{n|r + \beta|}$ exist, where $n = 1, 2, \dots$, $r = 0, \pm 1, \pm 2, \dots$, and $\beta$ depends on the trap parameters and sets the ion motional frequency \cite{Zhao2002}. 
Operation at these frequencies can be expected to be problematic as strong heating of the ion will be present, which we have observed.
This problem is easily solved because the width and strength of these resonances decrease with order, making them typically very narrow (typically 1~kHz) and difficult to excite. Small adjustments to the ion's motional frequency can thus shift the location of the parametric resonance sufficiently far outside the frequency of interest.

\subsection{Appendix I: Implementation of Quantum Transducer}

Following quantization of the electromagnetic field, the QVSA interaction Hamiltonian shown in Eq.~\ref{Hamil} can be expressed as $a b_d b_q^+ + \textrm{h.c}$, where $a$ is the  annihilation operator for a motional phonon; $b_d$ is the annihilation operator for a photon of the dipole field;
$b_q$ is the annihilation operator for a photon of the quadrupole field.

Assume that we want to do transduction from mechanical phonon to MW photon. 
Say from $\ket{n_a,n_d,n_q}=\sum_{n}{P(n)} \ket{1,n,0}$ to $\ket{n_a,n_d,n_q}=\sum_{n}{P'(n)}\ket{0,n-1,1}$.
Ions can be prepared to the $\ket{n=1}$ Fock state by applying a blue sideband pulse following sideband cooling.
A $\textrm{TE}_{21}$ mode microwave cavity, connected to a dilution refrigerator, in order to keep the occupation number of the quadrupole mode near zero, and resonant at the quadrupole frequency, is constructed around a surface ion trap. The dipole tone can then be applied classically via the surface trap electrode or a microwave horn.
This setup allows a single motional phonon to be converted into a single microwave photon, as well as the reverse process.
Furthermore, the number of photons generated into the cavity can be controlled by the Fock state occupation of the motional phonon.

\begin{table}
 \caption{Table of parameters for Figure 2}
  \centering
  \begin{tabular}{|c|c|c|c|c|}
    \toprule
    Subfigure & $\kappa V_d$ & $V_q$ & Total Time & Pulse Shaping?\\
    \midrule
    (a) & (0.239 $\pm$ 0.002) mV & (8.5 $\pm$ 0.1) mV& 1ms & Yes\\
    \midrule
    (b) & (0.489 $\pm$ 0.006) mV & (2.88 $\pm$ 0.03) V & 1ms & No\\
    \midrule
    (c) & (42 $\pm$ 3) mV & (3.8 $\pm$ 0.2) V& 1ms & No\\
    \midrule
    (d) & (0.226 $\pm$ 0.002) mV& (9.17 $\pm$ 0.01) mV & 1ms & Yes\\
    \midrule
    (e) & (2.20 $\pm$ 0.02) mV & (0.515 $\pm$ 0.005) V & 1ms & No\\
    \midrule
    (f) & (0.31 $\pm$ 0.02) V & (0.84 $\pm$ 0.05) V& 1ms & No\\
    \midrule
    (g) & See Figure & (8.62 $\pm$ 0.09) mV & 1ms & Yes\\
    \midrule
    (h) & See Figure & (1.36 $\pm$ 0.02) V & 1ms & No\\
    \midrule
    (i) & See Figure & (3.9 $\pm$ 0.3) V & 1ms & No\\
    \bottomrule
    \end{tabular}
    \label{tab:table2}
\end{table}

 \begin{table}
     \centering
    
     \begin{tabular*}{\linewidth}{@{\extracolsep{\fill}} ccccccccccc}
     \hline
         
          Carrier frequency (MHz)& 0.1&0.6&3&30&100&  200 & 350  &600 &800 &1000 \\
           \shortstack{Electric field sensitivity \\ (mV/m/$\sqrt{\textrm{Hz}}$)} &8.3(1.1)&2.1(3)&2.2(4)&8.1(1.7)&66(14) & 4.4(8) $\times 10^2$ & 1.3(3)$\times 10^3$  &2.5(5)$\times 10^3$& 3.6(8)$\times 10^3$ &1.0(2)$\times 10^4$\\
          Phase resolution (mrad)&-&-&9.4(7)&9.9(1.2)&15.7(2.5)&8.6(1.3)&22.8(10.2)&9.3(1.7)&9.6(2.9)&11.7(2.2)\\
    \hline
     \end{tabular*}
     \caption{The list of measurements shown in Fig.~\ref{fig:state_of_art_comp}. "-" means not applicable. All values listed in the table are derived from an Allan deviation following the same procedure used in Fig.~\ref{fig:below_sql}.}
     \label{table1}
 \end{table}